\documentclass[12pt,nofootinbib,notitlepage,preprint,aps,
  pra,
  amsmath,amssymb,longbibliography]{revtex4-2}
\usepackage{ulem}
\usepackage[varg]{txfonts}
\usepackage{bm}
\usepackage{setspace}
\usepackage{xcolor}
\usepackage[left]{lineno}

\PassOptionsToPackage{hyphens}{url}
\usepackage[bookmarks=true,        
     pdfnewwindow=true,      
     colorlinks=true,    
     linkcolor=blue,     
     citecolor=blue,     
     filecolor=blue,  
     urlcolor=blue,     
     final=true,
 ]{hyperref}
\usepackage{color,graphicx,amsmath,amssymb,fancyhdr}

\usepackage{footnote}

\makeatletter
\newcommand\footnoteref[1]{\protected@xdef\@thefnmark{\ref{#1}}\@footnotemark}
\makeatother
\usepackage{etoolbox}
\usepackage{footmisc}
\makeatletter
\patchcmd{\frontmatter@abstract@produce}
  {\vskip200\p@\@plus1fil
   \penalty-200\relax
   \vskip-200\p@\@plus-1fil}
  {}
  {}
  {}
\makeatother

\usepackage{siunitx}
\usepackage{threeparttable}
\newcommand{\Nuu}{\mathrm{Nu}}

\newcommand{\grouprow}[1]{\multicolumn{10}{@{}l}{\itshape #1}\\[1pt]}
\usepackage{booktabs}
\usepackage[export]{adjustbox}   

\begin{document}
\singlespacing
\renewcommand{\abstractname}{}
\title{Transport of Magnetic Fields by Thermohaline Convection in Crystallizing White Dwarfs}

\author{Nicolas Frazao Fernandes}
\affiliation{\rm TAPIR, California Institute of Technology, Pasadena, CA 91125, USA}
\author{J. R. Fuentes}
\affiliation{\rm TAPIR, California Institute of Technology, Pasadena, CA 91125, USA}
\author{Jim Fuller}
\affiliation{\rm TAPIR, California Institute of Technology, Pasadena, CA 91125, USA}

\begin{abstract}
The origins of strong magnetic fields observed in many white dwarfs are uncertain. Recent observations show that such fields are more common for white dwarfs with crystallized cores. Crystallization creates a destabilizing composition gradient that drives thermohaline convection in the star's fluid envelope, which may play an important role in magnetic field emergence. We perform 3D magnetohydrodynamical simulations of thermohaline convection with magnetic fields of various strengths and orientations, using the Boussinesq approximation in periodic Cartesian boxes. Our simulations suggest that thermohaline convection can generate magnetic fields when the magnetic Prandtl number is large (i.e., low magnetic diffusivity), but they are likely too weak to account for those observed in magnetic white dwarfs. However, the simulations indicate that strong magnetic fields can be advected outwards by the thermohaline convection without being destroyed, depending on the net magnetic flux. Hence, crystallization-induced thermohaline mixing may transport magnetic fields initially trapped in white dwarf interiors to their outer layers, allowing them to be observed. However, additional simulations using more realistic geometry and magnetic field configurations will be needed to confirm this possibility.

\end{abstract}


\maketitle

\section{Introduction} \label{sec:Introduction}

A substantial fraction of white dwarfs (WDs) have strong magnetic fields ($B \gtrsim 10^6 \, {\rm G}$) at their surfaces, but the origins of these magnetic fields are poorly understood (see \cite{Ferrario2022} for a review). Recent observations \cite{Bagnulo2021} show that the fraction of strongly magnetized WDs increases sharply at cool temperatures, soon after the cores of the WDs begin to crystallize \citep{Bagnulo2022}. It had already been suggested that crystallization powers a magnetic dynamo \cite{Isern2017} above the crystallizing core via crystallization-driven convection. The convection would be driven by a destabilizing composition gradient because the solid inner core is preferentially oxygen-rich (more dense), leaving behind a carbon-rich fluid layer above it (less dense), which can buoyantly rise into the heavier primordial carbon-oxygen fluid above it. \citet{Ginzburg2022} argued that the convective motions are too slow to produce strong magnetic fields unless the field strength can be increased by the WD's rotation.

However, \citet{Fuentes2023} and \citet{Montgomery_Dunlap2024} (see also \citet{Mochkovitch1983}) showed that stable thermal stratification of the WD interior prevents ordinary convection from occurring. Instead, the destabilizing composition gradient and the stabilizing entropy gradient cause thermohaline convection to occur. Thermohaline convection is enabled by thermal diffusion and grows on small length scales (or long time scales) compared to ordinary convection. \citet{Fuentes2023} estimate fluid velocities of only $v_{\rm therm} \sim 10^{-3} \, {\rm cm/s}$ depending on the influence of rotation. If the dynamo-generated magnetic field is limited to a value near equipartition, i.e., $B^2 \sim 4 \pi \rho v_{\rm therm}^2$, the expected magnetic fields are $B \sim 3 \, {\rm G}$, far too small to explain the observed values of highly magnetized WDs. It thus becomes very challenging to attribute strong magnetic fields to crystallization-driven dynamos. 

A possible savior for the crystallization-driven dynamo is that the initial stages of crystallization can induce genuine compositionally-driven convection \citep{Fuentes2024}, possibly creating velocities $v \sim 30 \, {\rm m/s}$ and magnetic fields of $B \sim 10^6 \, {\rm G}$. However, this magnetic field would be trapped in the deep core of the WD, and models predict that its strength at the WD surface would be greatly reduced to $B \lesssim 10^5 \, {\rm G}$, too low to account for observed field strengths of $B \gtrsim 10^6 \, {\rm G}$. 

It is also possible that many WDs are born with strong magnetic fields confined to their inner layers, undetectable at their surfaces. Indeed, asteroseismology shows that roughly 50\% of WD-producing stars with $M \gtrsim 1.5 \, M_\odot$ have strong core magnetic fields on the early red giant branch \citep{Fuller2015,Stello2016}. These magnetic fields may arise from a dynamo in the convective cores of main sequence stars. If their magnetic flux is conserved into later stages, the magnetic fields are expected to have $B \gtrsim 10^7 \, {\rm G}$ in the WD stage, but will be restricted to mass coordinates well below the surface for WDs with $M \lesssim 0.7 \, M_\odot$ \citep{Cantiello2016}. In these low-mass WDs, the magnetic diffusion time between the core and surface is more than $10$ Gyr \citep{Blatman2023}, so the magnetic fields would be trapped beneath the surface, unobservable. However, for relatively massive WDs ($M \gtrsim 0.65 \, M_\odot$), \citet{Camisassa2024} showed that the diffusion of such magnetic fields from WD interiors may be able to explain the onset of strong magnetic fields at WD surfaces, in addition to their field strengths \citep{CastroTapia2026}. In this model, it would merely be a coincidence that WDs become strongly magnetic about the same time as the core crystallization.

Another possibility, which is the motivating idea behind this work, is that thermohaline motions induced by core crystallization can help advect pre-existing core magnetic fields to the surfaces of WDs. Thermohaline motions would not be the source of the magnetic energy (overcoming the energetic obstacle discussed above), but they would help advect strong fields to the surface on short time scales (overcoming the diffusion time obstacle discussed above). However, it is not clear whether thermohaline motions would effectively advect a core magnetic field upwards, or whether they would instead act as an effective turbulent diffusivity that destroys the core magnetic fields. The nature of thermohaline motions will also be substantially altered by magnetic fields as shown in the works by \citet{Harrington2019,Fraser:2024}. 

The purpose of this work is to better understand these issues and to determine whether it is possible for thermohaline motions to advect magnetic fields outward in WDs without destroying them. To do this, we perform MHD simulations of thermohaline instability for various initial magnetic field strengths and geometries. Section 2 describes our simulation setup, Section 3 analyzes the results, and we discuss implications for magnetic WDs in Section 4 before concluding in Section 5.

\section{Numerical Simulations}

\subsection{Thermohaline Model}
\begin{figure*}
    \centering
\includegraphics[width=1\textwidth]{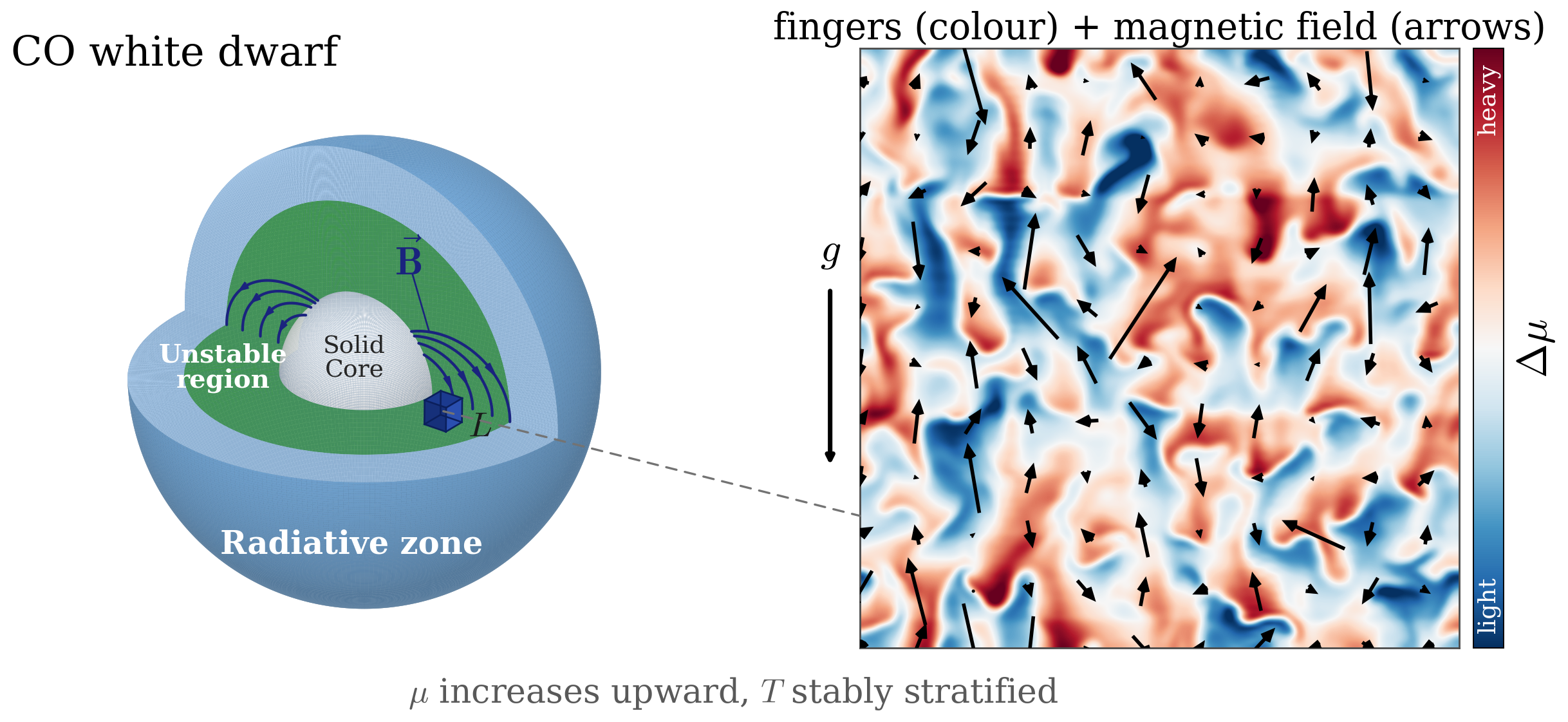}
    \caption{Left: a diagram of a crystallizing C/O white dwarf with an oxygen-rich solid crystalline core surrounded by a thermohaline layer in green. Our simulation domain spans a small portion of the thermohaline layer, with side length $L$ comparable to 100 $d$ where $d$ is the expected turbulent finger width. The right panel illustrates the fingering motions caused by thermohaline convection. Arrows show magnetic field vectors and colors show the molecular weight perturbation.}
    \label{fig:modeldiagram}
\end{figure*}

We investigate the interaction between thermohaline convection and magnetic fields with three-dimensional magnetohydrodynamical simulations. We focus on a local patch of the star near the crystallizing core (see Figure~\ref{fig:modeldiagram}) and use Cartesian geometry for our simulations. In doing so, we adopt the Boussinesq approximation of the fluid equations \citep{Spiegel_Veronis_1960}, i.e., we assume incompressibility, $\nabla \cdot \bm{v} =0$. Density perturbations arise through a linearized equation of state depending on temperature, $T$, and composition, $C$,
\begin{align}
\rho = \rho_0(1 - \alpha T + \beta C),
\end{align}
where $\rho_0$ is the mean density, while $\alpha$ and $\beta$ are the coefficients of thermal expansion and compositional contraction, respectively, both assumed to be positive constants. In our problem, we choose $C$ to represent the concentration of heavy elements. This approximation is appropriate for subsonic flows whose characteristic vertical length scale is much smaller than the local pressure scale height.

Our approximations are well justified near the crystallizing core because the characteristic velocity and length scale of thermohaline fingers are relatively small, with $v_{\rm th}< 1$ cm/s , and $L\sim 10$--$100\,d$, where $d\sim 10~\mathrm{cm}$ \citep[e.g.,][]{Montgomery_Dunlap2024}. Here,

\begin{align}
d = \left(\dfrac{\nu \kappa_T}{N^2_T}\right)^{1/4},
\end{align}
is the expected width of the turbulent finger and corresponds to the length scale of the fastest growing modes \citep[e.g.,][]{Harrington2019,Fraser:2024} , $\nu$ and $\kappa_T$ are the kinematic viscosity and thermal diffusivity, respectively, and
 
 \begin{align}
 N^2_T = \alpha g \left(\dfrac{dT_{\rm ad}}{dz} - \dfrac{dT_{0}}{dz}\right),
 \end{align}
is the square of the thermal contribution to the buoyancy frequency, where $g$ is the gravitational acceleration, $dT_0/dz$ is the background temperature gradient, and $dT_{\rm ad}/dz$ is the adiabatic temperature gradient. The compositional contribution to the buoyancy frequency is
\begin{align}
N^2_C = -\beta g \frac{dC_0}{dz} \, .
\end{align}

We choose background temperature and composition profiles such that both $T_0(z)$ and $C_0(z)$ vary linearly with $z$. The thermal stratification is stabilizing (sub-adiabatic), while the compositional stratification is destabilizing. The relative importance of these two contributions is quantified by the stability ratio, $R_0 $, and the Lewis number, $\mathrm{Le}$, defined respectively as

\begin{align}
    R_0 = \dfrac{|N^2_T|}{|N^2_C|}, \quad \mathrm{Le} = \dfrac{\kappa_T}{\kappa_C}, \label{eq:R0_Le}
\end{align}
where $\kappa_C$ is the chemical diffusivity. For $R_0  < 1$, the fluid is unstable to overturning convection due to strong compositional buoyancy. For $1 < R_0 < \mathrm{Le}$ the fluid is unstable to thermohaline convection, while for $R_0  > \mathrm{Le}$, the fluid is stabilized by the strong thermal buoyancy \citep[see, e.g.,][]{Garaud2021}. In crystallizing white dwarfs, the system typically evolves from $R_0  < 1$ at the onset of crystallization to  $R_0  > \mathrm{Le}$ as composition gradients are progressively eroded by mixing \citep{Fuentes2023, Castro-Tapia2024}. However, since we focus on how magnetic fields interact with thermohaline fingers, we restrict our simulations to the thermohaline regime $1 < R_0 < \mathrm{Le}$.

To investigate how thermohaline convection interacts with a pre-existing magnetic field, we consider several initial magnetic field configurations. In the first set of simulations, the magnetic field is spatially localized within part of the domain,

\begin{align}
 \bm{B}(x,y,z,t=0) = -\dfrac{ 1 - \tanh(\epsilon z) }{2}  \sin \left(\dfrac{2 \pi z}{L} \right) \bm{\hat{e}}_x, \label{half_horizontal_initB}
\end{align}
where $L$ is the vertical size of the domain and $\epsilon$ controls the sharpness of the transition. This configuration corresponds to a horizontal sinusoidal field in the $x$ direction, occupying approximately half of the domain in $z$, with a negligible magnetic field in the remaining region. We refer to this as the half-horizontal initial condition.

We also perform simulations initialized with a magnetic field spanning the whole domain, such that the net flux is zero

\begin{align}
 \bm{B}(x,y,z,t=0) =  \sin \left(\dfrac{2 \pi z}{L} \right) \bm{\hat{e}}_x, \label{sinusoidal_initB}
\end{align}
which we refer to as the whole-sinusoidal initial condition.

To capture multiple finger-like eddies in the thermohaline region, we set the horizontal and vertical size of the domain to $L = 100 d$. In addition, we impose periodic boundary conditions in all directions. This choice eliminates boundary effects and ensures that the magnetic flux through any plane is a conserved quantity, which constrains the possible evolution of the magnetic field.

\subsection{Fluid Equations and Dimensionless Parameters}
\label{Boussinesqequations}

We present the magnetohydrodynamic fluid equations in dimensionless form, using the width of the finger, $d$, and the thermal diffusion time across it, $d^2/\kappa_T$, as units of length and time, respectively. The units of temperature and composition are $[T] = d  \left|\frac{\partial T_0}{\partial z} - \frac{\partial T_{\rm ad}}{\partial z}\right|$ and $[C] = \frac{\alpha}{\beta}\,[T]$, while the magnetic field has units of $[B] = B_0$, the magnitude of the initial field at maximum amplitude. Under this choice, the dimensionless equations are

\begin{gather}
\frac{D\bm{v}}{Dt}
=-\nabla p
+\Pr\nabla^2 \bm{v}
+H_B(\nabla\times \bm{B})\times \bm{B}
+\Pr(T'-C')\bm{e}_z, \label{eq:momentum}\\
\frac{DT'}{Dt} + v_z
=\nabla^2 T', \label{eq:T}\\
\frac{DC'}{Dt} + \frac{v_z}{R_0}
= \mathrm{Le}^{-1}\nabla^2 C', \label{eq:C}\\
\frac{\partial \bm{B}}{\partial t}
=\nabla\times(\bm{v}\times \bm{B})
+\dfrac{\mathrm{Pr}}{\mathrm{Pm}}\nabla^2 \bm{B}, \label{eq:B}\\
\nabla\cdot \bm{v}=0, \\
\nabla\cdot \bm{B}=0, \label{eq:divB}
\end{gather}
where $D/Dt = \partial_t + \bm{v}\cdot \nabla$ is the material derivative, and $T'$ and $C'$ are perturbations about mean background states, $T_0(z)$ and $C_0(z)$.

The set of Equations~\eqref{eq:momentum}--\eqref{eq:divB} is governed by five dimensionless parameters:

\begin{gather}
\mathrm{Pr} = \dfrac{\nu}{\kappa_T}, \quad \mathrm{Pm} = \dfrac{\nu}{\eta}, \quad \mathrm{Le} = \dfrac{\kappa_T}{\kappa_C}, \\
H_B =  \dfrac{B_0^2 d^2}{\rho_0 \mu_0 \kappa_T^2}, \quad R_0 = \dfrac{|N^2_T|}{|N^2_C|},
\end{gather}
where $\mu_0$ is the magnetic permeability
of vacuum. Here, $\mathrm{Pr}$, $\mathrm{Pm}$, and $H_B$, are the Prandtl number, magnetic Prandtl number, and Chandrasekhar number\footnote{We define the Chandrasekhar number using the thermal diffusion time across a finger width $d$ instead of the viscous diffusion timescale.}, respectively, while $\mathrm{Le}$ and $R_0$ are the Lewis number and stability ratio, defined previously (Eq. \ref{eq:R0_Le}). These parameters can be expressed more concisely in terms of characteristic timescales. For example,

\begin{gather}
\mathrm{Pr} = \dfrac{\tau_{\kappa}}{\tau_\nu}, \quad \mathrm{Pm} = \dfrac{\tau_\eta}{\tau_\nu},\quad \mathrm{Le} = \dfrac{\tau_C}{\tau_{\kappa}}, \quad
H_B =  \dfrac{u^2_A}{u^2_\kappa}~,
\end{gather}
where $\tau_\kappa$, $\tau_\nu$, $\tau_\eta$, and $\tau_C$, are the thermal, viscous, magnetic, and chemical diffusion times in $d$, respectively, while $u_A = B_0/\sqrt{\rho_0 \mu_0}$ and $u_\kappa = \kappa_T/d$ are the Alfv\'en and thermal diffusion speeds, respectively. Although realistic stellar parameters are far beyond current computational capabilities for direct numerical simulations, we guide our choice of dimensionless parameters using cooling models of WDs computed with MESA  (see Appendix \ref{MESA_models}). In particular, we fix $\mathrm{Pr}  = \mathrm{Le}^{-1} = 0.1$, $\mathrm{Pm} = 1$ or $\mathrm{Pm} = 10$, and $R_0  = 3$, while varying $H_B $ in the range 0.01--100, in order to preserve the same ordering of diffusivities expected in the thermohaline region of crystallizing WDs ($\kappa_C \ll \nu \sim \eta \ \ll \kappa_T$). This ensures that the simulations remain in the same qualitative dynamical regime, with the only major deviation being that our $\mathrm{Le}$ should be much larger. 

\subsection{Quantities for Analysis}

We characterize the solutions of our simulations using volume-averaged quantities of several properties of the flow. We first define the volume-averaged kinetic and magnetic energies as

\begin{equation}
    \langle E_K \rangle \equiv  \frac{1}{V}\int \frac{v^2}{2}\,dV~, \qquad
     \langle E_B \rangle  \equiv  \frac{1}{V}\int H_B\frac{B^2}{2}\,dV~.
\end{equation}

The corresponding volume-averaged vertical fluxes of composition and temperature are

\begin{equation}
     \langle F_C \rangle  \equiv  \frac{1}{V}\int (v_z C)\,dV, \qquad
     \langle F_T \rangle  \equiv  \frac{1}{V}\int (v_z T)\,dV.
\end{equation}
We use these fluxes to define the thermal and compositional Nusselt numbers,

\begin{equation}
     \langle \mathrm{Nu}_T \rangle  \equiv 1 + \langle v_z T  \rangle, \qquad
     \langle \mathrm{Nu}_C \rangle \equiv 1 + R_0 \mathrm{Le} \langle v_z C  \rangle~.
    \label{volumeNC}
\end{equation}

We additionally characterize the relative importance of the magnetic and kinetic energies using the equipartition ratio,

\begin{equation}
     \Lambda  \equiv \frac{E_B}{E_K}.
    \label{volumeLambda}
\end{equation}

Time averages are computed according to

\begin{equation}
\overline{Q} \equiv \dfrac{1}{\Delta t}\int_{t_0}^{t_0+\Delta t} Q\,dt.
\end{equation}
Since we focus on the quasi-stationary state, time averages are taken over the final $20\%$ of each simulation. In the following, time-averaged volume-average quantities are written without brackets or overbars for simplicity.

\subsection{Numerical Methods}

We time-evolve the equations \eqref{eq:momentum}--\eqref{eq:divB} using the Dedalus pseudospectral solver \citep{Burns2020}, version 3. All variables are expanded on the Fourier basis using 128 coefficients in each direction. In the appendix, we also include simulations with double or triple $L_z$ in which we scale the number of Fourier modes in the z direction accordingly. For time-stepping, we use a Runge-Kutta second order two stage scheme \citep[RK222,][]{Ascher1997}, where the linear and nonlinear terms are treated implicitly and explicitly, respectively. To ensure numerical stability, the size of the time steps is set by the Courant–Friedrichs–Lewy (CFL) condition, using a safety factor of 0.2. Both the flow speed and Alfv\'en speed are used as characteristic velocities of the system. To prevent aliasing errors, we apply the "3/2 rule'' in all directions when evaluating nonlinear terms. To start the simulations, we add random-noise temperature and composition perturbations sampled from a normal distribution with a magnitude of $10^{-3}$.

\section{Results} \label{sec:Results}

Here, we present the results of direct numerical simulations of equations \eqref{eq:momentum}--\eqref{eq:divB}. As expected from linear theory \cite{Charbonnel2007}, our simulations show fingering instability with a short exponential growth phase. Saturation is determined by visual inspection of the time series of the volume-averaged kinetic and magnetic energies, $E_K$ and $E_B$, respectively. We find that flow saturation or lack thereof is highly dependent on the initial magnetic field configuration and the Chandrasekhar number $H_B  \propto B_0^2$ that sets the initial magnetic field strength. Therefore, we separate the results into different subsections for different simulation setups (see Table \ref{tab:summarytable}).
The first two sets of "half-horizontal" simulations (A1-A5) and (B1-B5) have a magnetic field with the initial condition of Eq. (\ref{half_horizontal_initB}), which entails a non-zero magnetic flux through the boundary. The B1-B5 runs suppress mean flows in the z-direction by suppressing $k_z=0$ modes (see Section \ref{subsec:halfsineconstrained}). The set of runs (C1-D5) have "whole-sinusoidal" initial condition with Eq. (\ref{sinusoidal_initB}), and therefore have zero initial magnetic flux. They differ only in that $\mathrm{Pm}=1$ for C runs and $\mathrm{Pm}=10$ for D runs.

\begin{table*}
  \centering
  \footnotesize
  \setlength{\tabcolsep}{5pt}
  \sisetup{round-mode=places, round-precision=2, group-digits=none}
  \begin{threeparttable}
    \caption{Summary of all simulations. $E_K$, $E_B$ and $v_{\rm rms}$ are volume
             averages further averaged in time over the last $20\%$ of each run
             (the quasi-stationary phase); runs that do not saturate are flagged
             in the text. Initial field geometries are given by
             Eq.~\eqref{half_horizontal_initB} (half-horizontal, non-zero net
             flux) and Eq.~\eqref{sinusoidal_initB} (whole-sinusoidal, zero net
             flux).}
    \label{tab:summarytable}
    \begin{tabular}{
    l
    S[table-format=3.2, round-mode=none]   
    S[table-format=1.2, round-mode=none]   
    S[table-format=2.0, round-mode=none]   
    c
    @{\hspace{1.5em}}
    S[table-format=2.5]                    
    S[table-format=1.2e-2, exponent-mode=scientific, print-zero-exponent] 
    S[table-format=1.3]                    
    S[table-format=3.2]                    
    S[table-format=4.2]                    
    }
      \toprule
      & \multicolumn{4}{c}{Input parameters}
      & \multicolumn{5}{c}{Time-averaged diagnostics} \\
      \cmidrule(lr){2-5} \cmidrule(l){6-10}
      Run & {$H_B $} & {$R_0 $} & {$\mathrm{Pm}$} & Box aspect
          & {$E_K$} & {$E_B$} & {$v_{\rm rms}$} & {$\Nuu_T$} & {$\Nuu_C$} \\
      \midrule
 
      \grouprow{(a) Half-horizontal, Eq.~\eqref{half_horizontal_initB} --
                unconstrained elevator modes}
      \addlinespace[1pt]
      A1 &  0.01 & 3 &  1 & 1:1:1 &  0.2298 & 7.52e-3 & 0.5556 & 1.6548 & 32.9996 \\
      A2 &   0.1 & 3 &  1 & 1:1:1 &  0.2900 & 6.918e-2 & 0.6458 & 1.9711 & 48.3523 \\
      A3 &     1 & 3 &  1 & 1:1:1 &  1.5625 & 1.964e0 & 1.638 & 10.606 & 428.3425 \\
      A4 &    10 & 3 &  1 & 1:1:1 &  23.0182 & 6.3964e1  & 6.6365 & 174.471 & 7217.248 \\
      A5 &   100 & 3 &  1 & 1:1:1 &  4.8754 & 8.4844e1  & 2.9284 & 46.3665 & 1896.579 \\
      \addlinespace[4pt]
 
      \grouprow{(b) Half-horizontal, Eq.~\eqref{half_horizontal_initB} --
                constrained elevator modes ($k_z=0$ frozen)}
      \addlinespace[1pt]
      B1 &  0.01 & 3 &  1 & 1:1:1 &  0.2026 & 6.5787e-3 & 0.51785 & 1.5538 & 28.432 \\
      B2 &   0.1 & 3 &  1 & 1:1:1 &  0.2411 & 5.884e-2 & 0.58375 & 1.7701 & 39.280 \\
      B3 &     1 & 3 &  1 & 1:1:1 &  0.4165 & 4.581e-1 & 0.8040 & 2.7848 & 90.274 \\
      B4 &    10 & 3 &  1 & 1:1:1 & 0.2629 & 8.293e-1  & 0.58018 & 2.0241 & 51.869 \\
      B5 &   100 & 3 &  1 & 1:1:1 &  0.5041 & 5.064e0  & 0.19883 & 1.1536 & 6.9802 \\
      \addlinespace[4pt]
 
      \grouprow{(c) Whole-sinusoidal, Eq.~\eqref{sinusoidal_initB} --
                zero net magnetic flux, low Pm}
      \addlinespace[1pt]
      C1 &  0.01 & 3 &  1 & 1:1:1 &  0.2242 & 1.361e-5 & 0.5447 & 1.6276 & 31.4173 \\
      C2 &   0.1 & 3 &  1 & 1:1:1 &  0.2266 & 4.290e-5 & 0.5480 & 1.6297 & 31.5283\\
      C3 &     1 & 3 &  1 & 1:1:1 &  0.2236 & 2.565e-4 & 0.5440 & 1.6092 &  30.9802\\
      C4 &    10 & 3 &  1 & 1:1:1 &  0.2277 & 3.792e-3 & 0.5512 & 1.6389 & 32.1166\\
      C5 &   100 & 3 &  1 & 1:1:1 &  0.2492 & 1.914e-2 & 0.5840 & 1.7634 & 38.1168\\

      \grouprow{(d) Whole-sinusoidal, Eq.~\eqref{sinusoidal_initB} --
                zero net magnetic flux, high Pm}
      \addlinespace[1pt]
      
      D1 &  0.01 & 3 & 10 & 1:1:1 & 0.3333 & 1.265e-1 &  0.7061 & 2.1958 & 58.8001 \\
      D2 &   0.1 & 3 & 10 & 1:1:1 & 0.3349 & 1.2831e-1 & 0.7127 & 2.2588 & 60.9500 \\
      D3 &     1 & 3 & 10 & 1:1:1 & 0.3452 & 1.4234e-1 & 0.7207 & 2.2605 & 61.9080 \\
      D4 &    10 & 3 & 10 & 1:1:1 & 0.7125 & 5.3966e-1 & 1.0811 & 4.4643 & 156.9558\\
      D5 & 100 & 3 & 10 & 1:1:1 & 1.8922 & 2.0584e0 & 1.8276 & 12.1079 & 478.1036\\
      \addlinespace[4pt]
 
      \grouprow{(e) Gaussian-noise seed field }
      \addlinespace[1pt]
      E1 & 10 & 1.45 & 10 & 1:1:1 & 2.0771 & 8.0298e-1 & 1.7117 & 10.1847 & 220.1453\\  
      E2 & 10 & 3.0 & 10 & 1:1:1 & 0.3228 & 1.1835e-1 & 0.6935 & 2.1563 & 56.9979 \\ 
      E3 & 10 & 5.0 & 10 & 1:1:1 &  0.0553 & 1.0126e-2& 0.2857 & 1.1871 & 14.4071 \\ 
      E4 & 10 & 7.0 & 10 & 1:1:1 & 0.0088 & 1.3072e-10 & 0.1126 & 1.0330 & 3.9265 \\
      E5 & 10 & 7.0 & 10 & 2:2:2 & 0.0088 & 7.6306e-09 & 0.1126& 1.0331 & 3.9342 \\
      \addlinespace[4pt]
 
      \grouprow{(f) Box-size study -- as (b) with $H_B =1$, elongated domains}
      \addlinespace[1pt]
      F1 & 1 & 3 & 1 & 1:1:2 & 1.193 & 1.754e0  & 1.4165 & 7.8712 & 317.6144 \\
      F2 & 1 & 3 & 1 & 1:1:3 & 1.787 & 2.858e0 & 1.7591 & 14.580 & 587.6621 \\
      \bottomrule
    \end{tabular}
 
    \begin{tablenotes}[flushleft]\footnotesize
      \item Every run uses $\mathrm{Le}=10$,
            $\mathrm{Pr}=0.1$, $128$ Fourier modes per direction and a
            horizontal domain with periodic boundaries in all
            directions. "Box'' gives $L_x:L_y:L_z$ in units of $100\,d$; the
            number of  modes is scaled with $L_x:L_y:L_z$ in each direction.
    \end{tablenotes}
  \end{threeparttable}
\end{table*}

\begin{figure*}
    \centering
    \setlength{\tabcolsep}{0pt}
    \begin{tabular}{@{}c@{\hspace{4pt}}c@{\hspace{2pt}}c@{}}
    & \hspace*{-8pt}Half-horizontal, unconstrained\hspace*{8pt} & \hspace*{-8pt}Half-horizontal, constrained\hspace*{8pt} \\[2pt]    \adjustbox{valign=c}{\rotatebox{90}{$t=0$}}&
      \includegraphics[width=0.5\textwidth,valign=c]{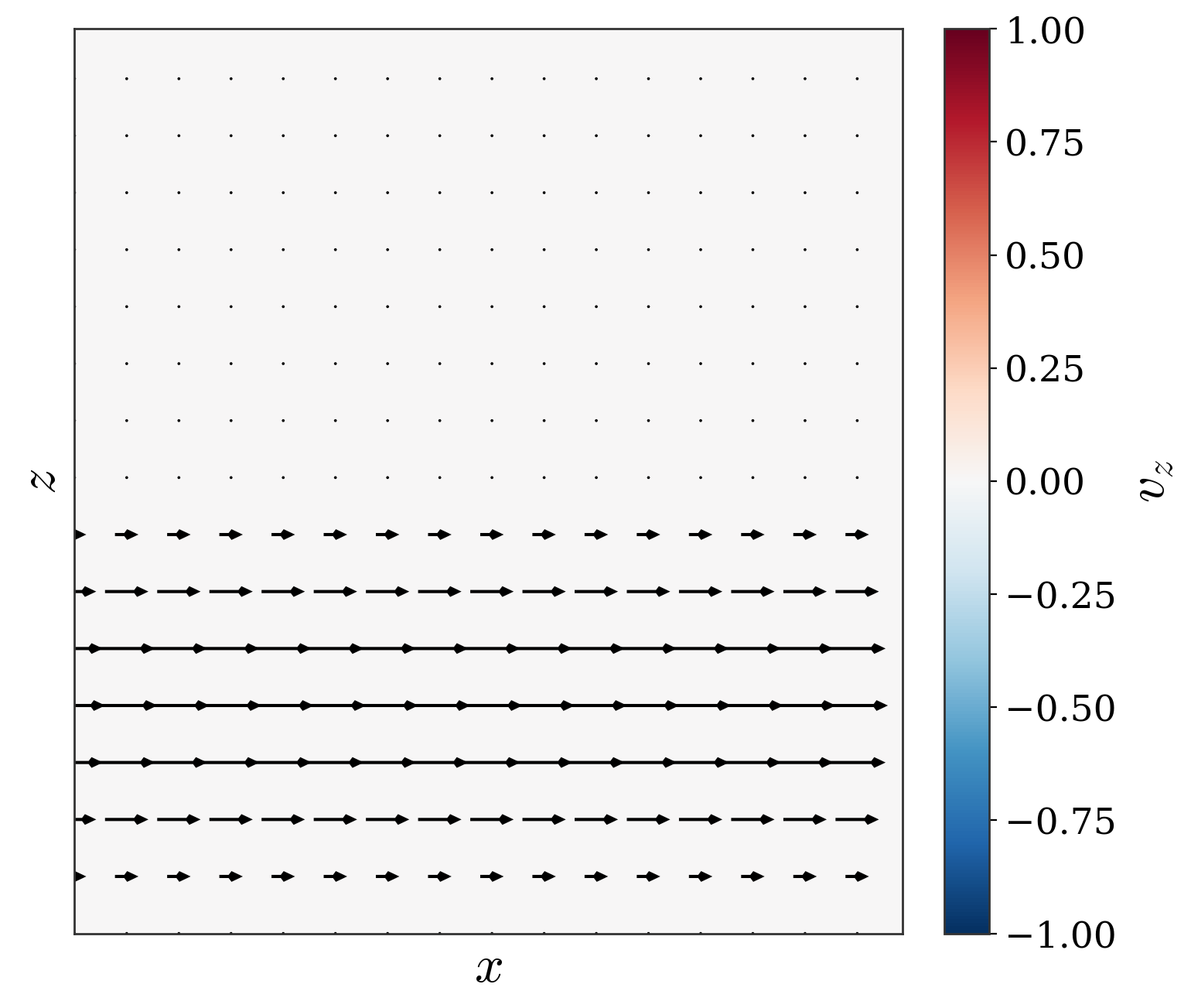} &
      \includegraphics[width=0.5\textwidth,valign=c]{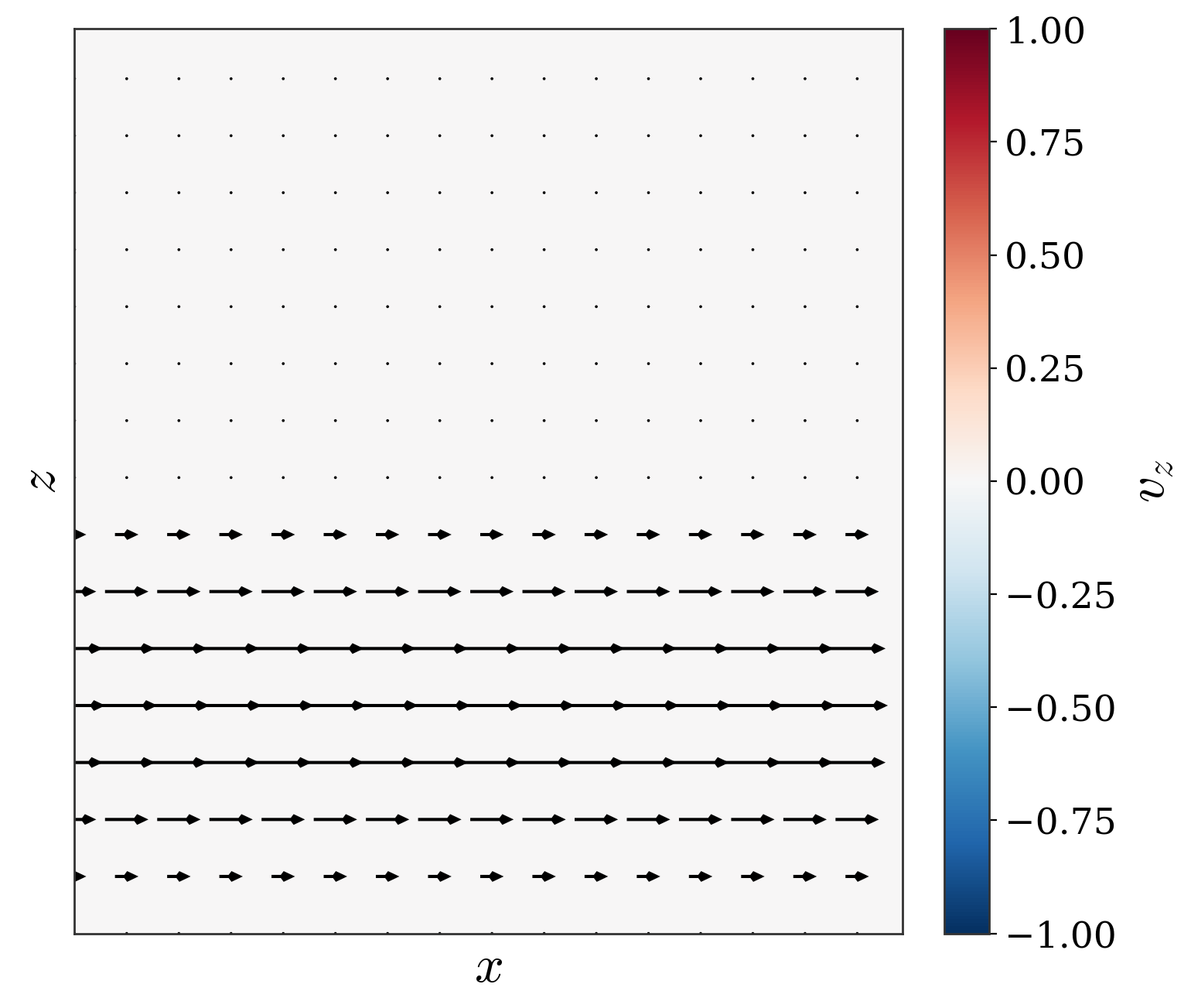} \\[2pt]
    \adjustbox{valign=c}{\rotatebox{90}{$t=1000$}}&
      \includegraphics[width=0.5\textwidth,valign=c]{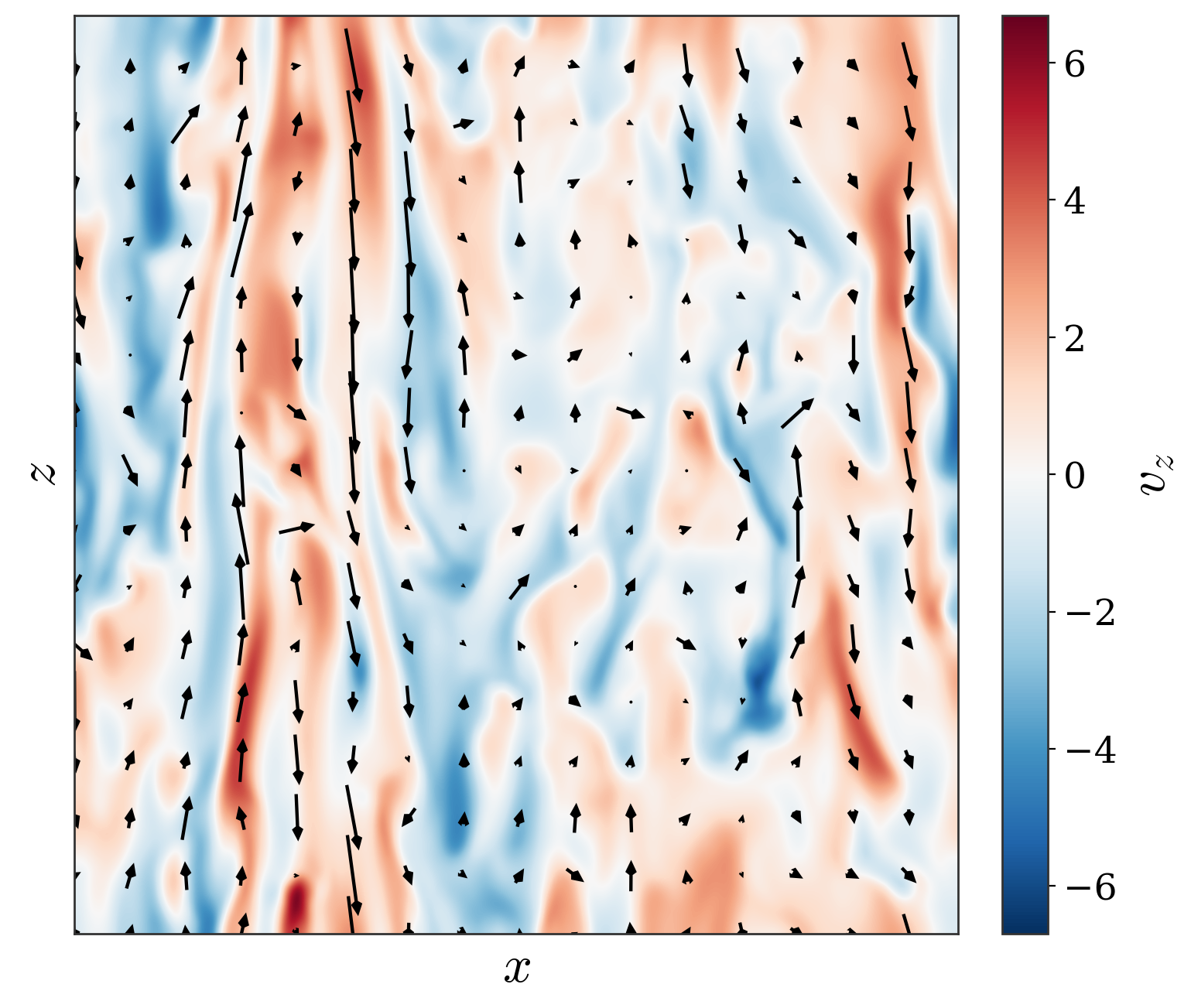} &
      \includegraphics[width=0.5\textwidth,valign=c]{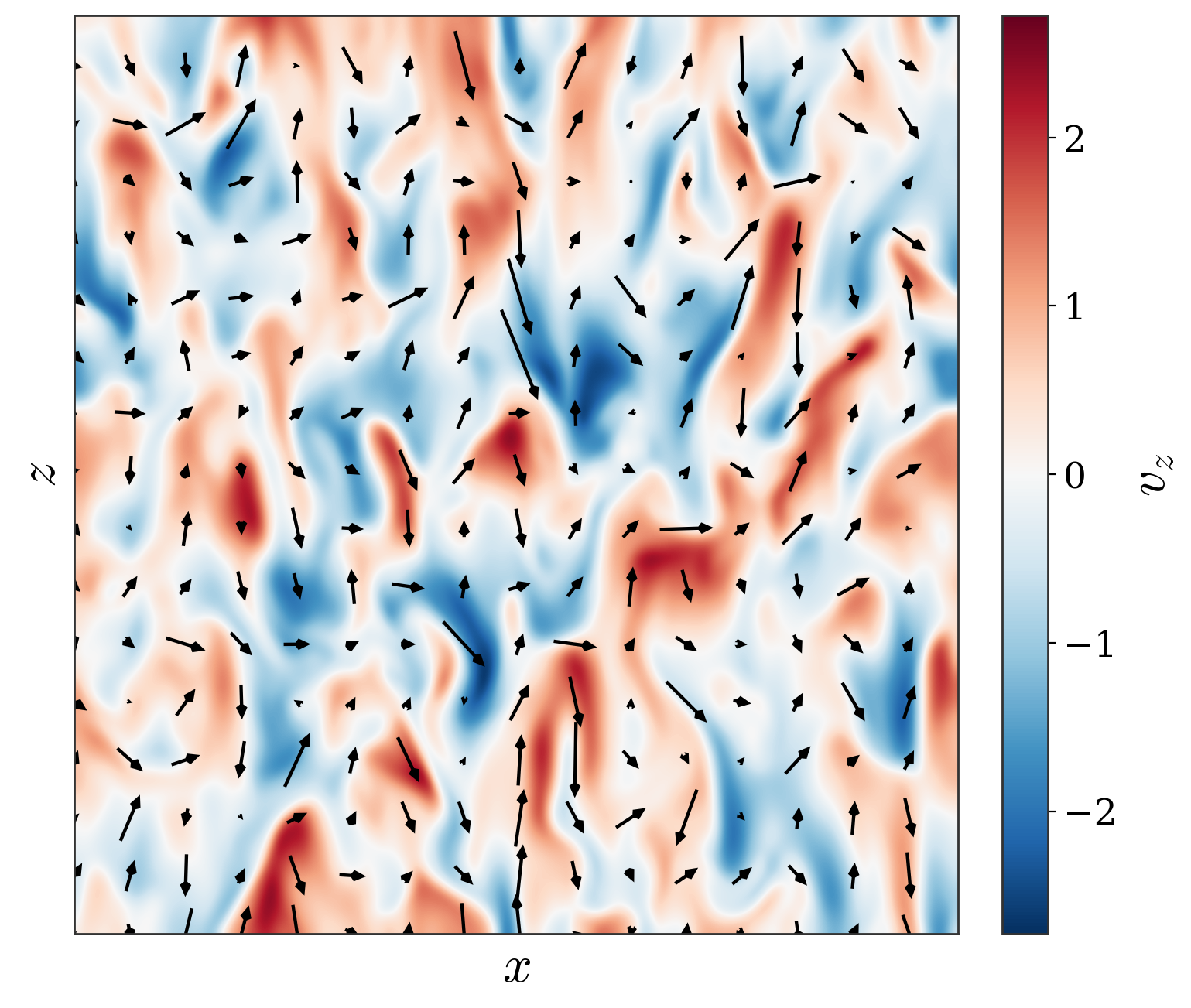}
    \end{tabular}
    \caption{Simulation results for run A3 (left column) and B3 (right column). The first row shows the initial conditions, both given by (Eq. \ref{half_horizontal_initB}), and the second row shows their final state at $t=1000$ with fully developed thermohaline convection. Color shading indicates the vertical velocity $v_z$ in the $xz$-plane at $y=0$. The arrows represent the relative magnitude and direction of magnetic field.  The right plots have constrained elevator-modes ($k_z=0$), which results in much different flow patterns, and smaller velocities.}
    \label{fig:elevator-comparison}
\end{figure*}

\begin{figure*}
  \centering
  \setlength{\tabcolsep}{0pt}
  \begin{tabular}{@{}c@{\hspace{4pt}}c@{\hspace{2pt}}c@{}}
      & Whole-sinusoidal, $\mathrm{Pm}=1$ & Whole-sinusoidal, $\mathrm{Pm}=10$ \\[2pt]
    \adjustbox{valign=c}{\rotatebox{90}{$t=0$}} &
      \includegraphics[width=0.5\textwidth,valign=c]{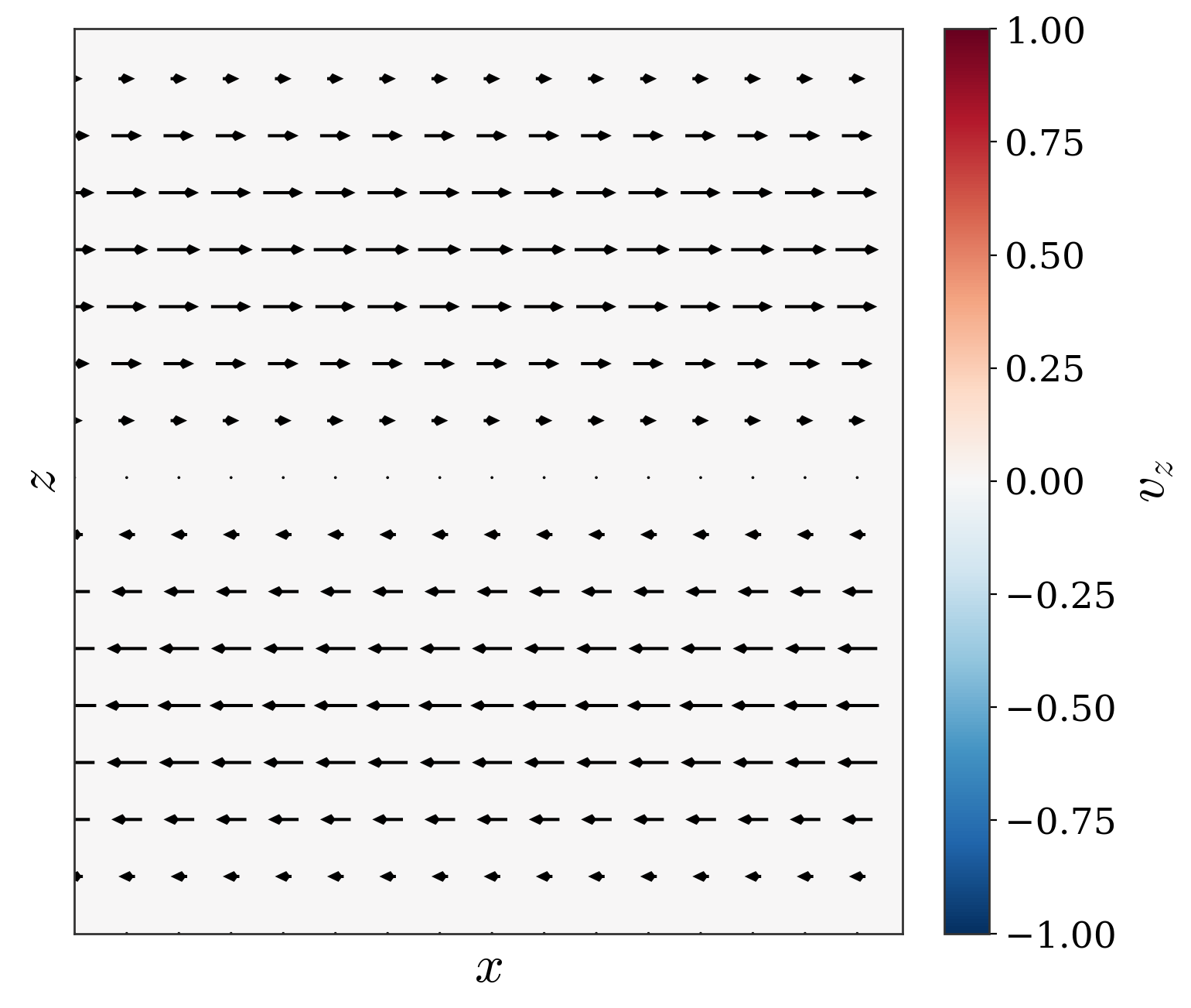} &
      \includegraphics[width=0.5\textwidth,valign=c]{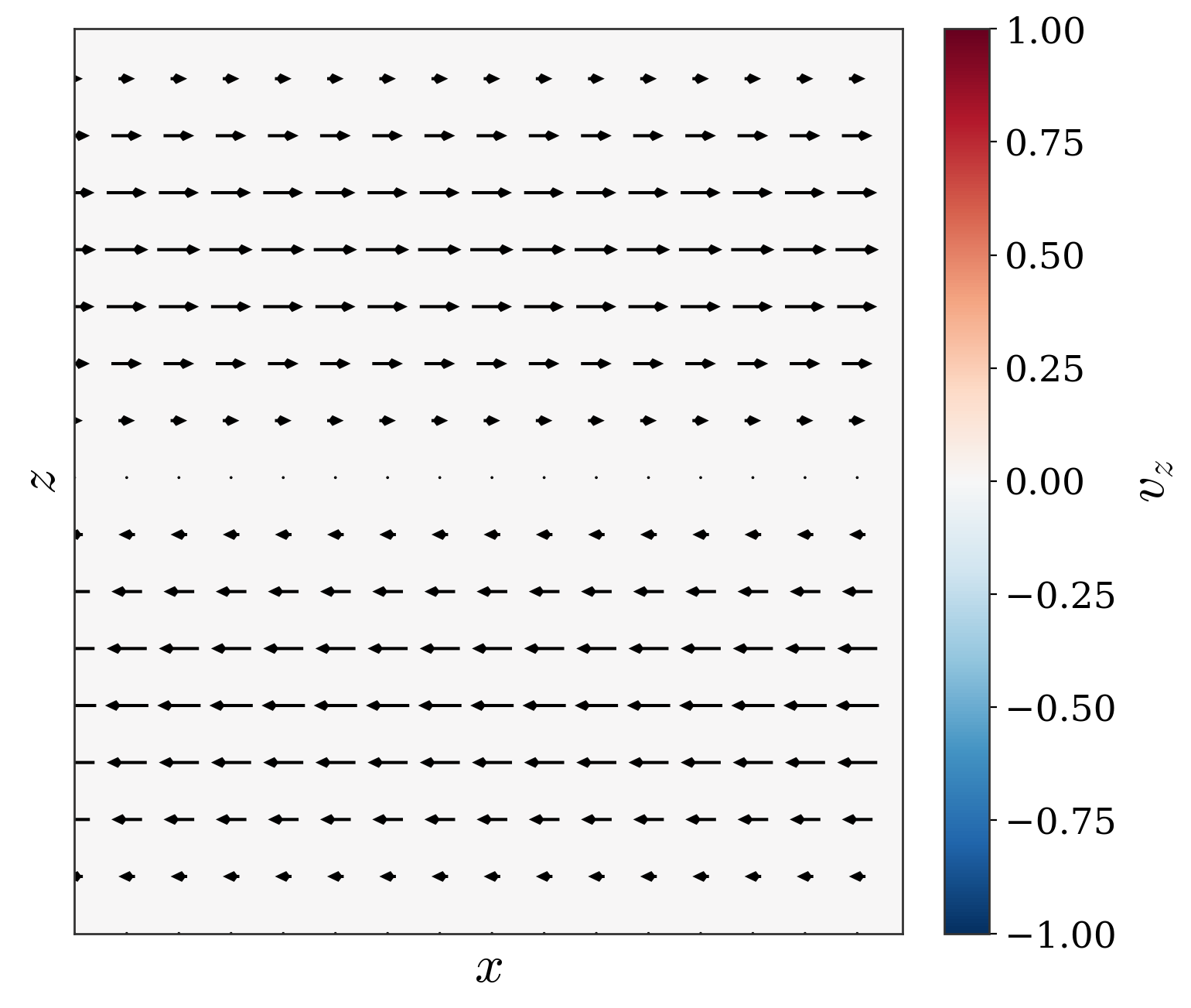} \\[4pt]
    \adjustbox{valign=c}{\rotatebox{90}{$t=1000$}} &
      \includegraphics[width=0.5\textwidth,valign=c]{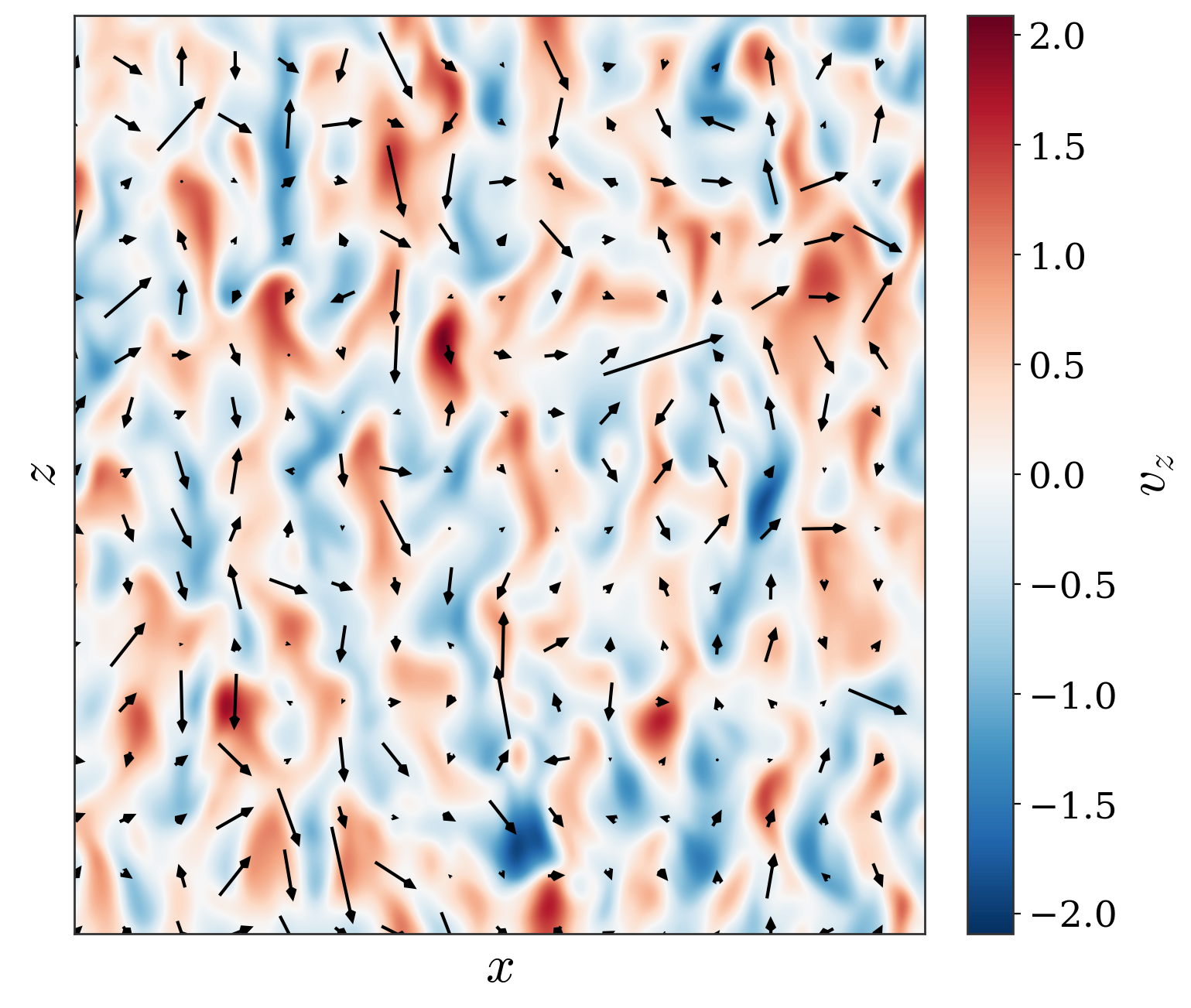} &
      \includegraphics[width=0.5\textwidth,valign=c]{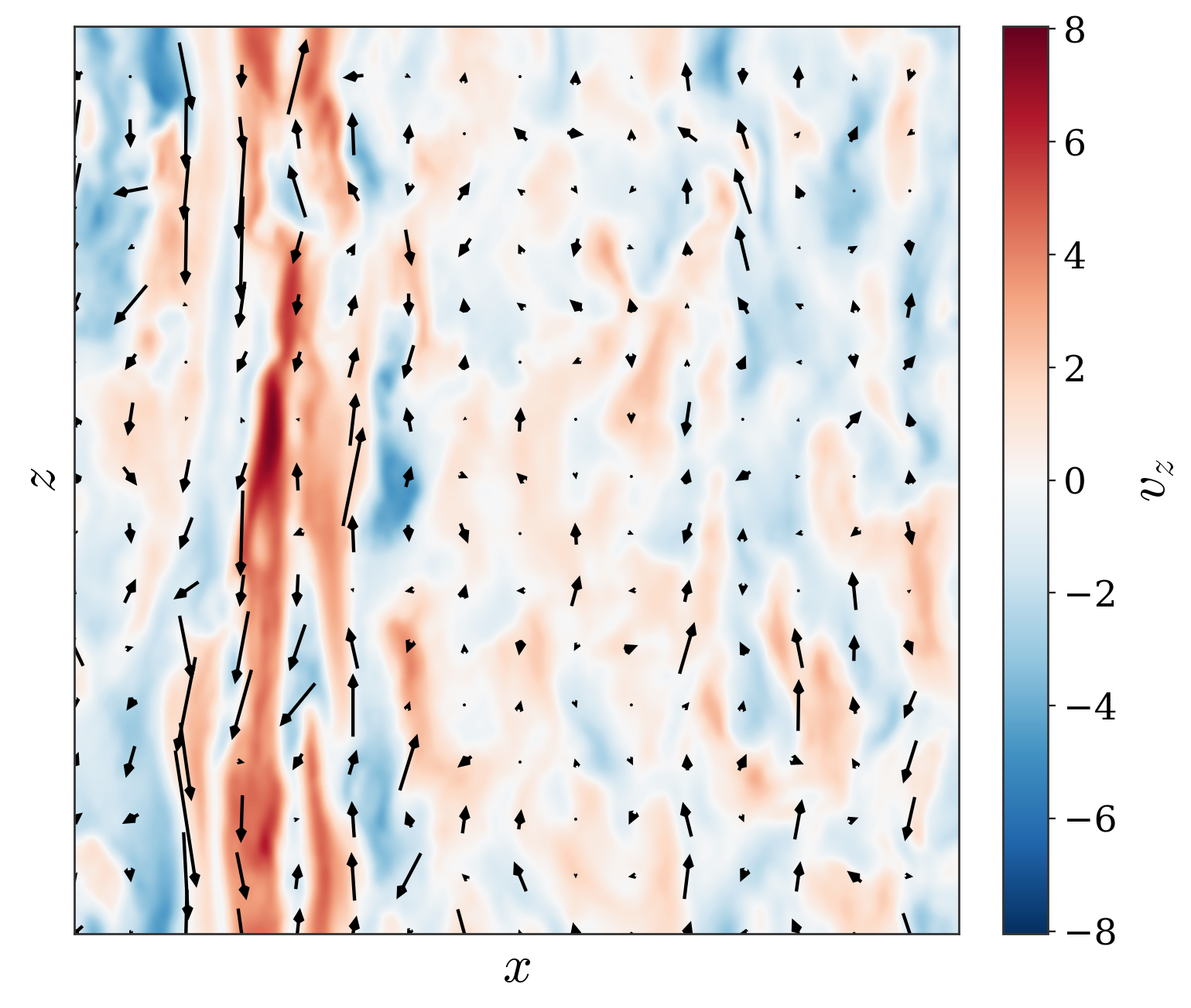}
  \end{tabular}
  \caption{Same as Figure (\ref{fig:elevator-comparison}), for simulations C5 and D5. 
  The snapshots in the left column have $H_B =100$ and $\mathrm{Pm}=1$ while the right has $H_B =100$ and $\mathrm{Pm}=10$.}
  \label{fig:zero-flux}
\end{figure*}

In all simulations, the magnetic field energy $E_B = H_B B^2 /2 $ initially decays slightly during the exponential growth phase. This behavior is consistent with the ohmic decay from magnetic diffusion. After this initial phase, the magnetic energy evolves differently in each type of simulation, which we discuss below.

\subsection{Half-horizontal initial condition}
\label{subsec:halfsine}

\begin{figure*}
    \centering
    \includegraphics[width=1\linewidth]{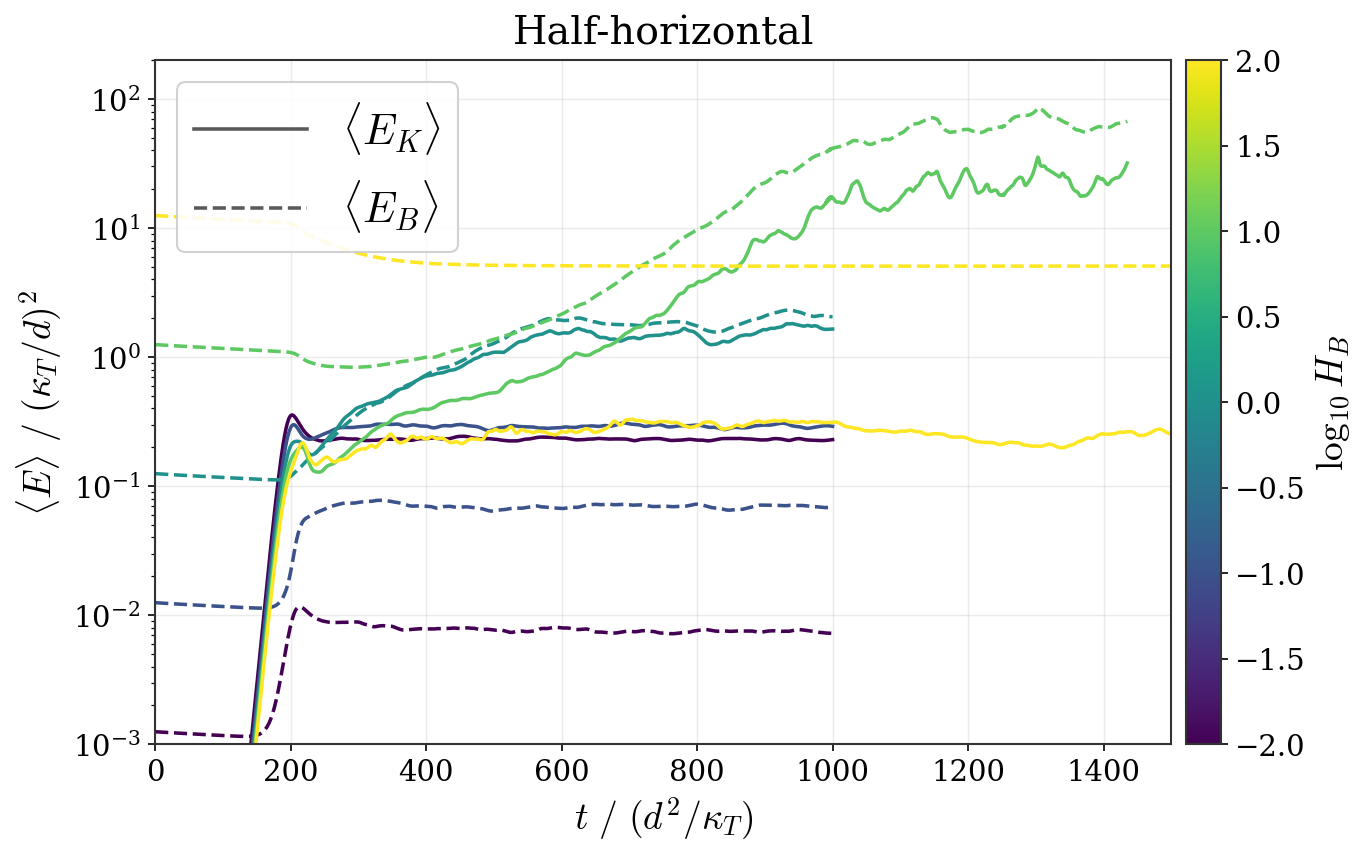}
    \caption{Time series of the volume-averaged kinetic energies ($E_K$, solid lines) and magnetic energies ($E_B$, dashed lines) for simulations A1-A5, with initial magnetic field given by Eq. \ref{half_horizontal_initB}. Simulations with $H_B  \geq 1$ do not reach a saturated steady-state. }
    \label{half_horizontal}
\end{figure*}

First, we will describe the simulations A1-A5 in Table \ref{tab:summarytable} with the initial condition (\ref{half_horizontal_initB}). Two xz-plane slices at y = 0 coordinate are shown in the left column of Figure \ref{fig:elevator-comparison} . At times $t=0$, magnetic field is initially non-zero and with net horizontal flux while vertical velocity is zero. The final saturated state at $t=1000$  has elongated finger-like structures that span the whole domain vertically. Additionally, the final state magnetic field arrows in the saturated state are predominantly aligned with the fingers, either up or down (conserving zero net flux in the z-direction). These features exist in the final state of all A1-A5 simulations. 

 Figure  \ref{half_horizontal} shows the volume-averaged energies as a function of time. The saturation of these simulations qualitatively changes between $H_B  < 1$ (A1 and A2) and $H_B  \geq 1$ (A3, A4, and A5). For \(H_B < 1\) we see an exponential increase in magnetic energy until the simulations reach the quasi-stationary state where the magnetic and kinetic energies become approximately constant with time. This saturated averaged magnetic energy is several times larger than the initial magnetic energy, but smaller than the kinetic energy. The so-called parasitic-saturation model for non-magnetized thermohaline convection proposed in previous works \citep{Brown2013,Harrington2019,Fraser:2024} explains the equilibrium for these models. The magnetic fields only have a minor effect on the outcome. 

The simulations with \(H_B  > 1\) do not reach a steady state. Analyzing the magnetic energy in each direction, we notice that the magnetic energy in the z-direction increases over time, while the field in the x-direction decreases over time. However, the magnetic energy in the x-direction has a minimum possible value set by the conservation of magnetic flux as shown in the Appendix \ref{fluxConservationProof}. For the half-horizontal field configuration (Eq.\ref{half_horizontal_initB}), the minimum magnetic energy is \(  E_B  \geq \frac{H_B}{2 \pi^2} \simeq 0.05 H_B \), while the initial magnetic energy is \(  E_B(t=0)  = 0.125 H_B \).

For the simulation with $H_B  = 10$ (A4), after the initial exponential growth phase for the linear instability (which ends at $t\approx 200$), we observed an ongoing and intermittent growth of the flow volume-averaged energies that persisted until the end of our simulations. We suspect this is because the parasitic shear modes that usually saturate thermohaline fingers are suppressed by the strong magnetic fields that develop. The z-component of the magnetic field in this simulation intermittently grows over time due to the combination of a minimum field in the $x-$direction (as discussed above) along with fast flows in the $z-$direction that stretch the $x-$component of the magnetic field into the $z-$direction.  These fast flows can develop due to our periodic boundaries, which permits $k_z=0$ "elevator modes" to grow without limit. The elevator modes should be the fastest growing modes of our system and they dominate at late times in these simulations. 

\begin{figure*}
    \centering
    \includegraphics[width=1\linewidth]{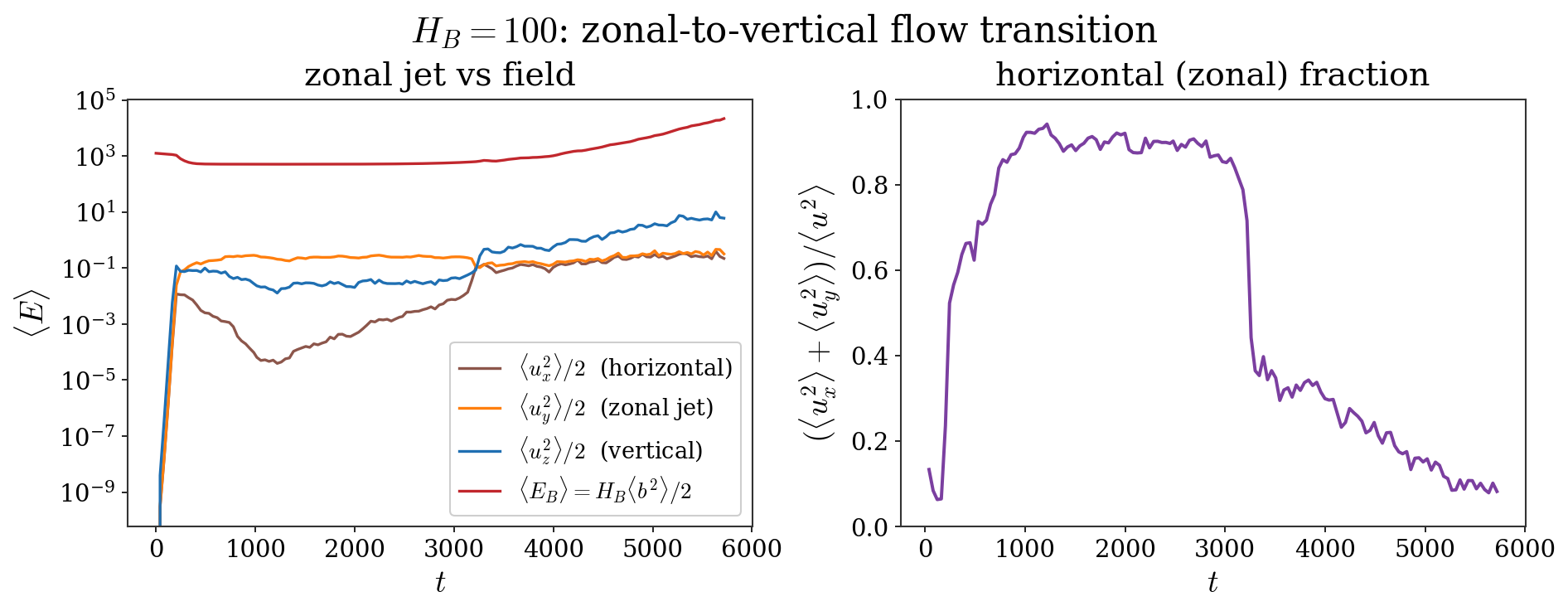}
    \caption{Simulation A5 with strong initial magnetic field, $H_B =100$. The left panel shows the volume-averaged energies, with the x-component of the kinetic energy (brown), the y-component kinetic energy (orange), the z-component kinetic energy (blue), and the magnetic energy (red). The right panel shows the total fraction of the kinetic energy in the horizontal directions. Notice that there is a transient characterized by the development of a y-direction zonal flow from $t \approx 200 -3200$. After this transient, the vertical kinetic energy and magnetic energy continue to grow, as in simulation A4 with $H_B =10$ (Figure \ref{half_horizontal}).}
    \label{fig:zonal}
\end{figure*}

For simulations A3 and A4, the magnetic fields act to stabilize the flow as also noted in \cite{Harrington2019,Fraser:2024} and prevent saturation in our simulations. However, in spherical geometry without triply periodic boundary conditions, these elevator modes with \(k_z = 0\) do not exist. We therefore believe that growth and saturation may proceed very differently in realistic spherical stars (see the next section).

Simulation A5, with the strongest initial magnetic fields ($H_B =100$) behaves very differently. At early times, this simulation has most of its kinetic energy distributed in horizontal zonal flows. These zonal flows disrupt the growth of thermohaline fingering motion in the vertical direction. While Figure \ref{half_horizontal} seems to indicate it has reached a steady state, we ran this simulation much longer (Figure \ref{fig:zonal}). By the end of the simulation, the zonal flows are beginning to dissipate, and motion in the vertical direction is beginning to increase, suggesting  that these zonal flows may be an initial transient phase. At the end of the simulation, the magnetic and $z$-component of the kinetic energy exhibit ongoing growth, similar to simulation A4 discussed above.

\subsection{Half-horizontal initial condition with constrained elevator modes}
\label{subsec:halfsineconstrained}

\begin{figure*}
    \centering
    \includegraphics[width=1\linewidth]{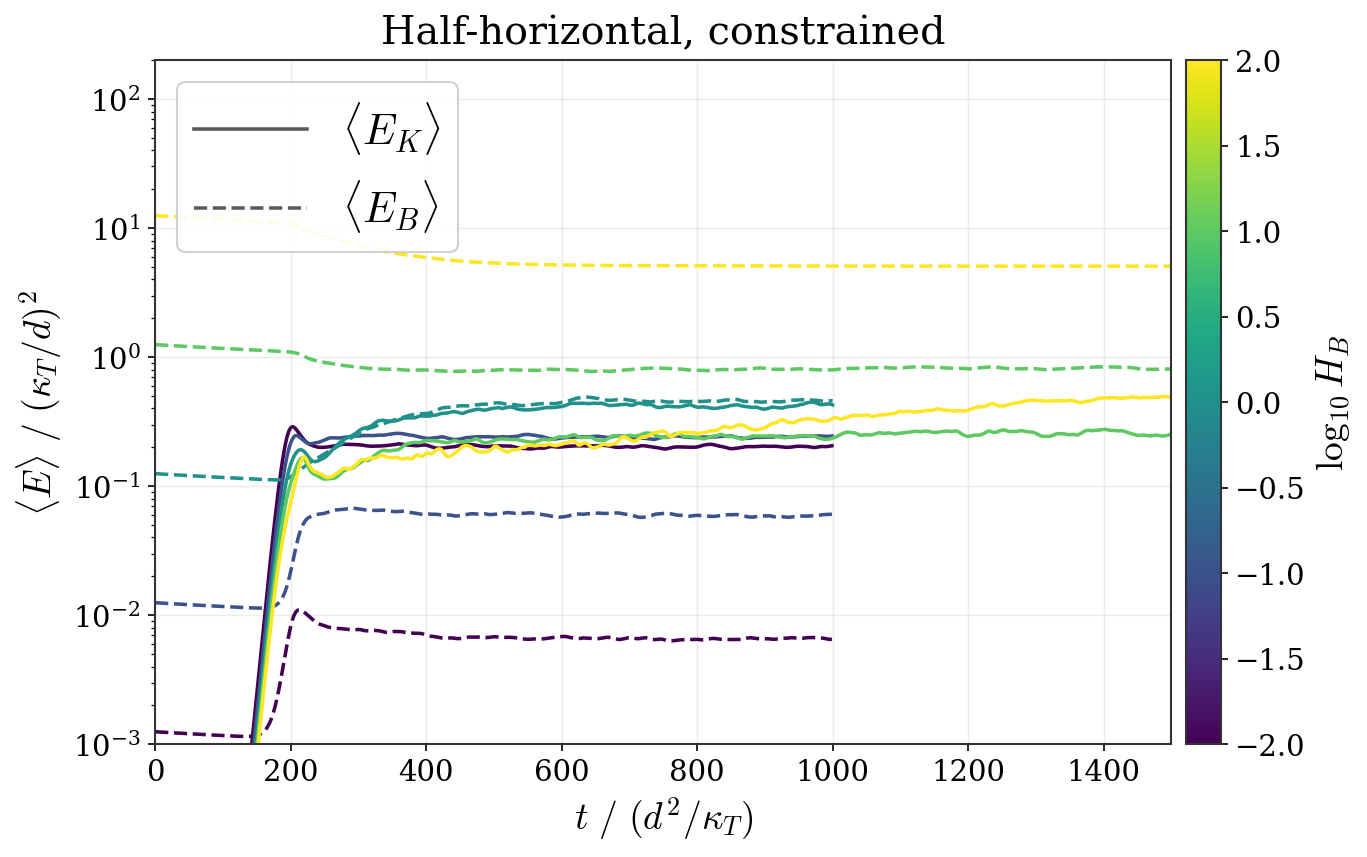}
    \caption{Time series of the volume-averaged kinetic energies ($\langle E_K \rangle$) and magnetic energies ($\langle E_B \rangle$) in constrained elevator mode simulations (B1-B5) with initial field given by Eq.  \protect(\ref{half_horizontal_initB}). These reach a steady state, apart from the simulation with $H_B  = 100$, due to development of transient zonal flows (see text).}
    \label{constrained_half_horizontal}
\end{figure*}

To suppress the contribution of the elevator modes, we have a family of simulations B1-B5 in which we changed our equations for the modes with \(k_z=0\) so that they remain constant and only the modes with \(k_z \neq 0\) evolve according to Equations~\eqref{eq:momentum}--\eqref{eq:divB}. The right column of  Figure \ref{fig:elevator-comparison}  has B3 slices at two different times as an example. Initially, at $t=0$, the magnetic field is also non-zero with net horizontal flux and no vertical velocity. The final saturated state at $t=1000$ is slightly different from that of the A1-A5 simulations. The finger-like structures are smaller and less coherent. Additionally, the final state magnetic field arrows have a tendency to be vertical, but they still continue to have a noticeable horizontal trend to the right.

Simulations B1-B4 all reach a quasi-stationary steady state as seen in Figure \ref{constrained_half_horizontal}. The case with the highest initial magnetic field (B5) with $H_B  = 100$  has a qualitatively different quasi-stationary equilibrium at late times that resembles the long transient phase of the unconstrained simulation A5. We suspect that it would eventually reach a steady state similar to simulations B1-B4 if run long enough. 

The saturated kinetic energies of simulations B1-B4 are quite similar, with little influence of the initial magnetic field. The late-time influence of the magnetic field is weak, even though an ordered magnetic field persists as required by flux conservation. The saturated magnetic energies of B1-B3 are higher than the initial field energy because of the generation of small-scale magnetic fields by thermohaline motions. The B4 simulation magnetic energy saturates at a lower value than the initial field energy, but higher than the lower bound from the flux conservation, likely because the ordered field in the $x$-direction is continually regenerating fields in the other directions due to thermohaline motions. 

Although these simulations (apart from the one with $H_B  = 100$) do reach a steady state, their properties depend on the size of the box (see Figure \ref{fig:boxsize} in Appendix \ref{convergence}).

\subsection{Whole-sinusoidal initial condition}
\label{subsec:fullsine}

\begin{figure*}
    \centering
    \includegraphics[width=1\linewidth]{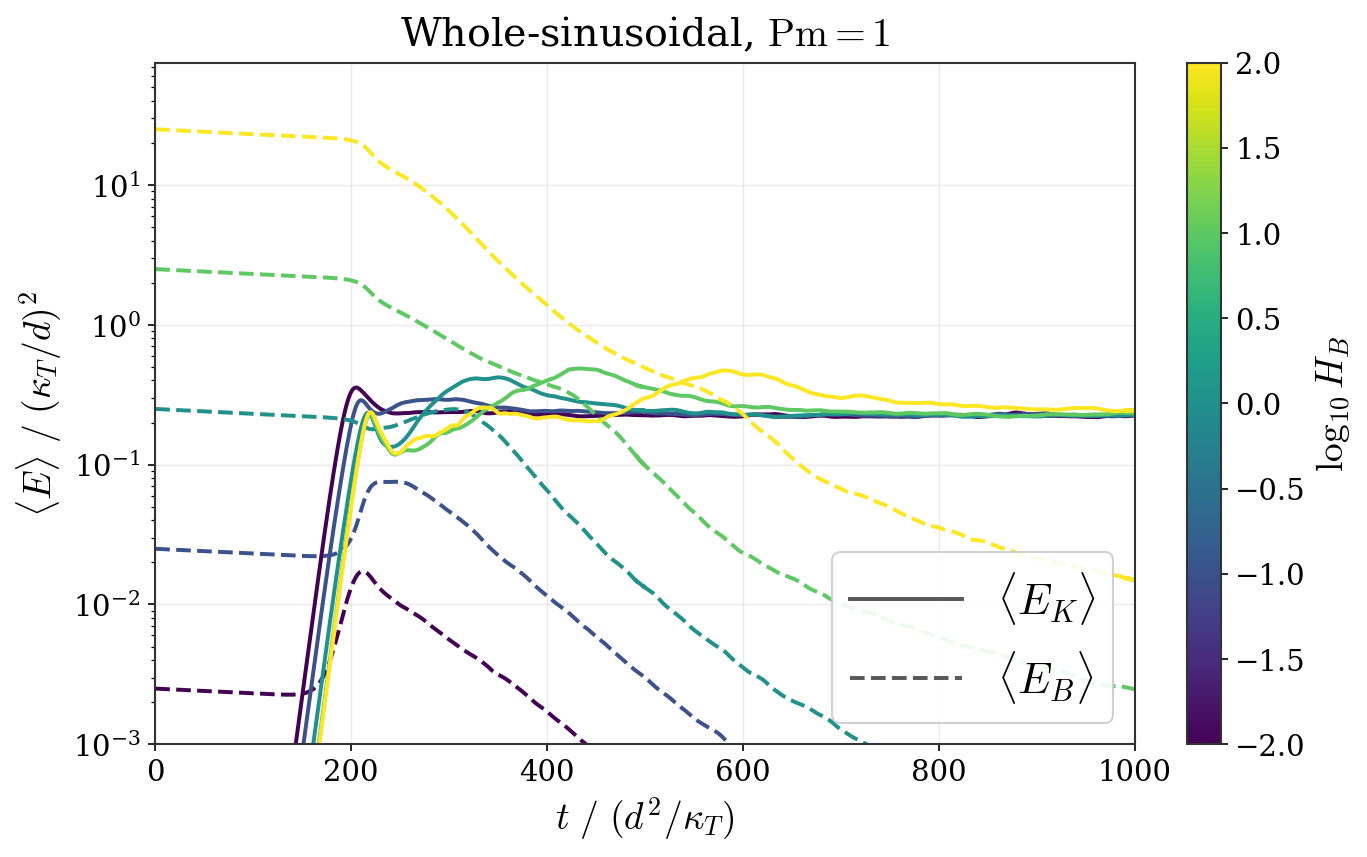}
    \caption{Time series of the volume-averaged kinetic energies $\langle E_K \rangle$ and magnetic energies $\langle E_B \rangle$ in whole-sinusoidal simulations (initial condition given by Eq. \ref{sinusoidal_initB}) and $\mathrm{Pm}=1$. These simulations have zero net flux, and the magnetic field decays in all cases, with no small-scale dynamo. The kinetic energies approach the pure hydrodynamical values.}
    \label{fig:fullsinepm1}
\end{figure*}

In simulations with initial magnetic fields consisting of a sinusoidal profile through the whole box (Eq. \ref{sinusoidal_initB}, simulations C1-D5), there is no net flux of the magnetic field through the boundaries. Therefore, there is no minimum mean magnetic field in any direction and the magnetic field can decay through magnetic reconnection. Figure \ref{fig:zero-flux} shows simulations C5 and D5 (left and right columns respectively) at $t=0$ (top), and at $t=1000$ (bottom). The simulation with $\mathrm{Pm}=1$ in the snapshot $ t=1000$ shows two distinct features: the field has no preferred direction and is nearly independent of the velocity field structure. On the other hand, the $\mathrm{Pm}=10$, $t=1000$ simulation shows coherent vertical fingers with vertically aligned magnetic fields. 

The magnetic field dissipates through turbulent magnetic reconnection in simulations C1-C5 (see the time series in Figure \ref{fig:fullsinepm1}), decaying towards values of zero.  The final quasi-stationary states with different $H_B $ have nearly identical kinetic energies and vanishing magnetic field energies.

\begin{figure*}
    \centering
    \includegraphics[width=1\linewidth]{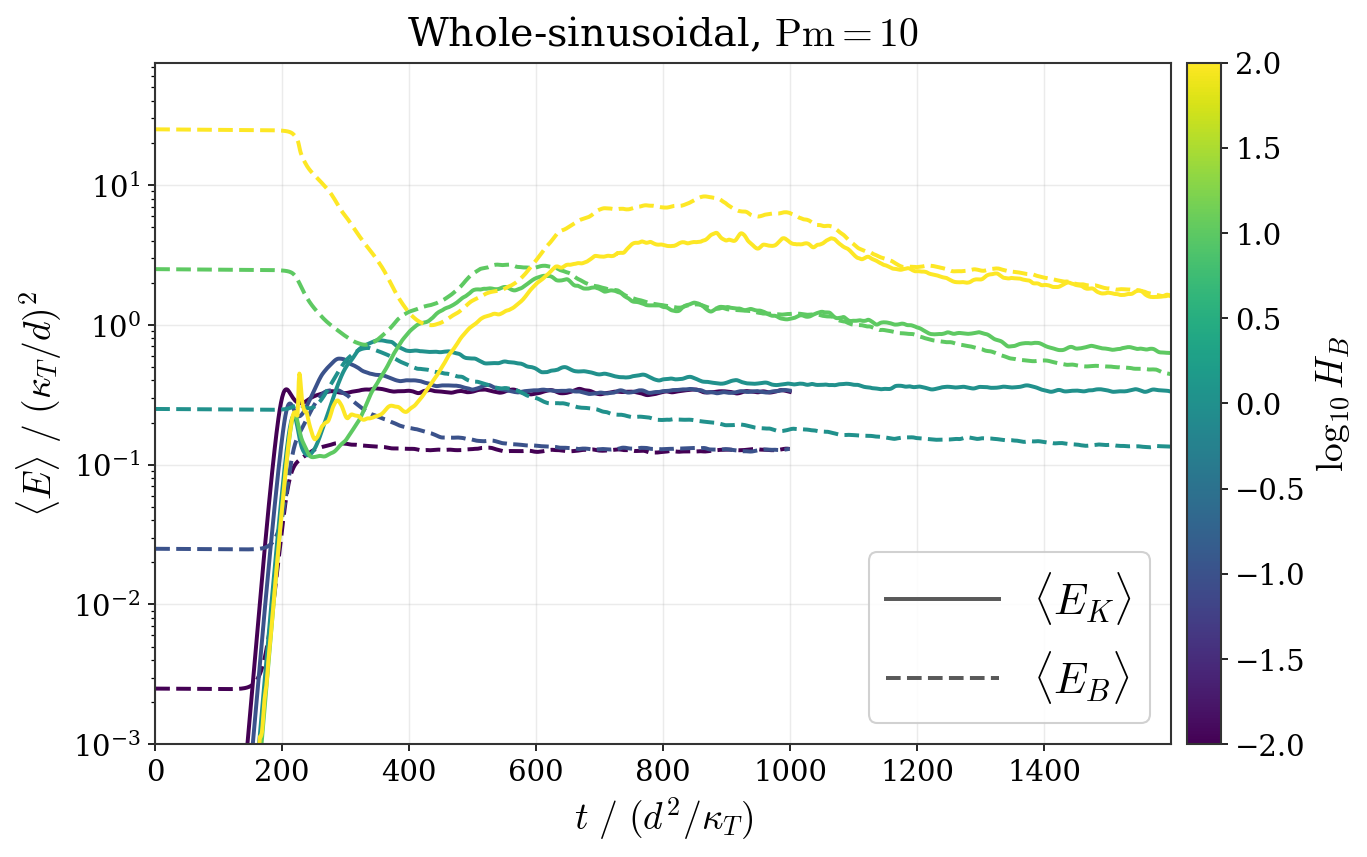}
     \caption{Time series of the volume-averaged kinetic energies $\langle E_K \rangle$ and magnetic energies $\langle E_B \rangle$ in whole-sinusoidal simulations with initial condition given by Eq.  \ref{sinusoidal_initB} and $\mathrm{Pm}=10$. These simulations exhibit small-scale dynamos that generate magnetic fields with energies a few times smaller than the kinetic energy.}
    \label{fig:pm10_timeseries}
\end{figure*}

For simulations D1-D5 with $\mathrm{Pm}  =10$, we found that the fingering convection induced a small scale dynamo that restored the field to energies a few times smaller than the kinetic energy in the saturated state (see Figure \ref{fig:pm10_timeseries}). These simulations exhibit a small-scale dynamo because the magnetic Reynolds number is larger, with $Re_m \equiv \frac{U L}{\eta} \approx \frac{v_{\rm rms} d}{\eta} $ , where $U$ is the characteristic speed of the flow (in our case the rms finger velocity $u_\text{rms}$) and $L$ (the finger width d) the characteristic length scale. The value of $Re_m$ is $\approx$10 times higher in the simulations with $\mathrm{Pm}=10$ compared to the simulations with $\mathrm{Pm}= 1$, with values of $O(10^2)$.

We note that simulations D4 and D5 in Figure \ref{fig:pm10_timeseries} have not reached steady state by t=1000. They appear to be decaying towards the same steady state reached by the other simulations, with $E_K \approx 0.3$ and $E_B \approx 0.1$. However, the large initial magnetic fields in these simulations temporarily allow for more vigorous thermohaline motions in which kinetic and magnetic energy are enhanced.

\begin{figure*}
    \centering
    \includegraphics[width=1\linewidth]{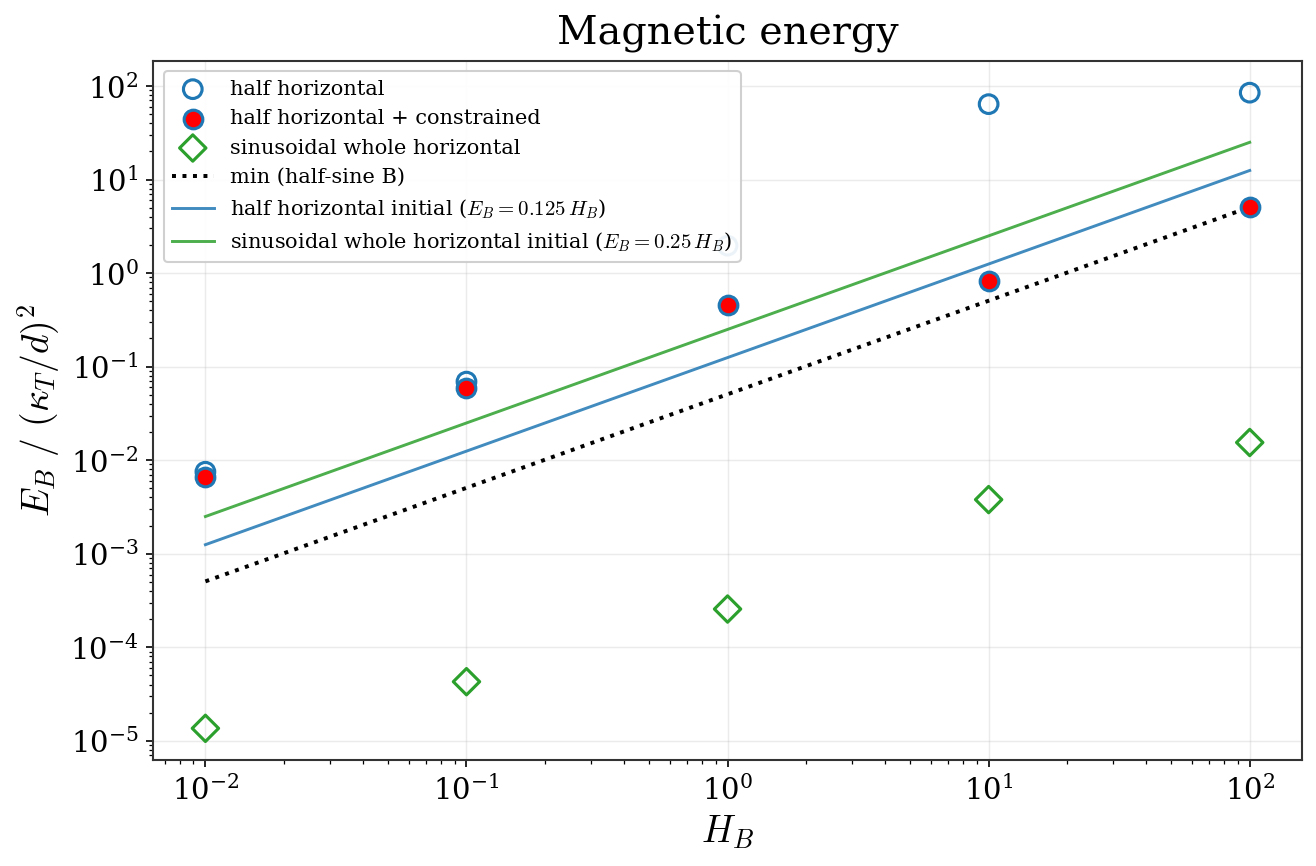}
    \caption{Final volume-averaged and time-averaged magnetic energy for our simulations A1-C5. For a half-horizontal initial magnetic field (with a finite magnetic flux), hollow blue circles are simulations with no constraint on the \(k_z = 0\) 'elevator' modes, while red-filled circles are simulations with constrained 'elevator' modes. Green diamonds are simulations with whole-sinusoidal initial magnetic field (zero magnetic flux). Solid lines show initial magnetic energies, and the dotted line shows the minimum possible field from flux conservation for the half-horizontal case.}
    \label{fig:BE}
\end{figure*}

\begin{figure*}
    \centering
    \includegraphics[width=1\linewidth]{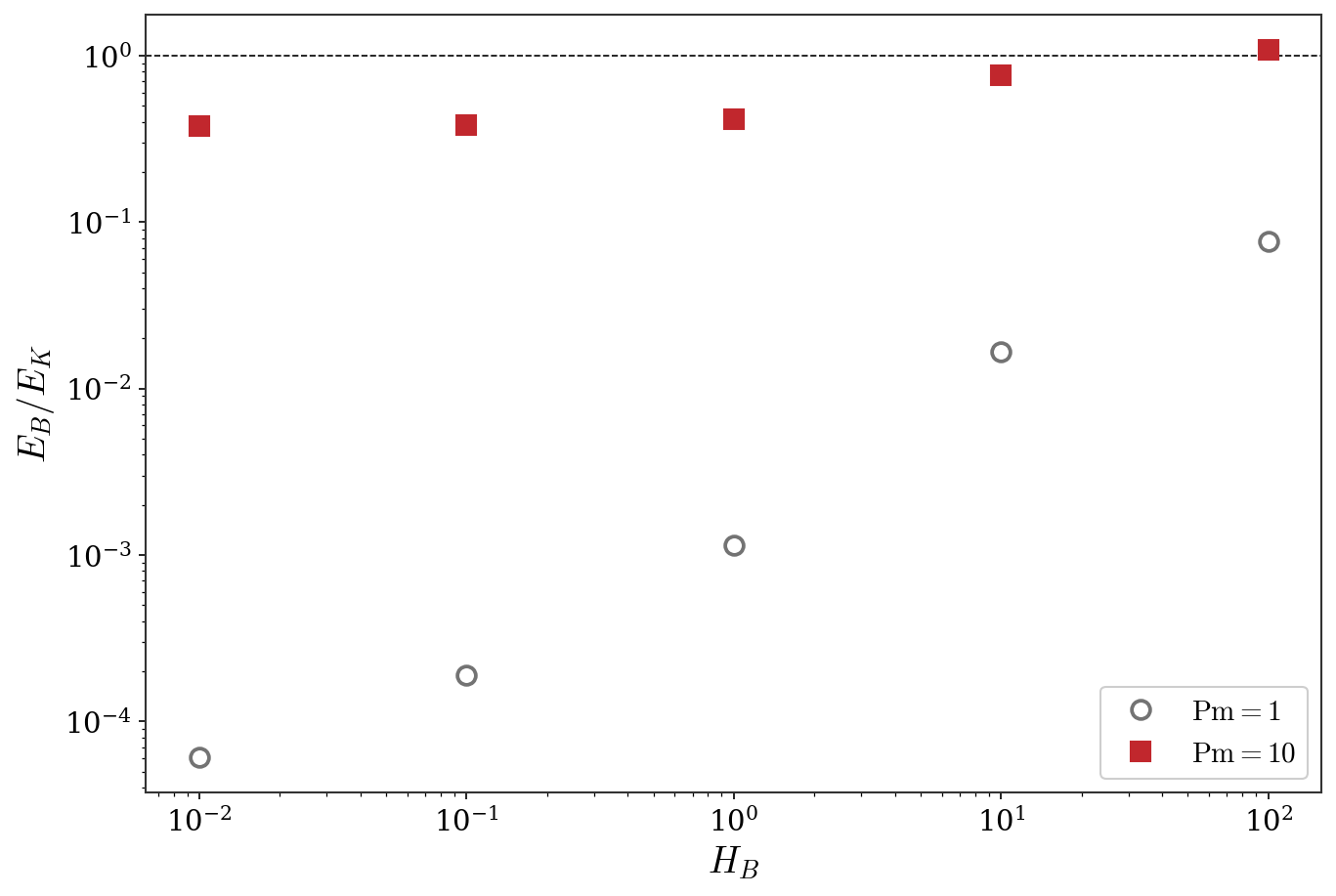}
    \caption{Comparison between the ratio of magnetic to kinetic energy, for runs with $\mathrm{Pm}  = 1$ and $\mathrm{Pm}=10$, and whole-sinusoidal initial magnetic field (Eq. \ref{sinusoidal_initB}) with zero net flux. $\mathrm{Pm}=10$ simulations result in dynamo-generated fields that saturate at near-equipartition values, while $\mathrm{Pm}=1$ simulation magnetic energies decay over time showing no signs of dynamo action.}
    \label{fig:pm10_equipartition}
\end{figure*}

\begin{figure*}
    \centering
    \includegraphics[width=1\linewidth]{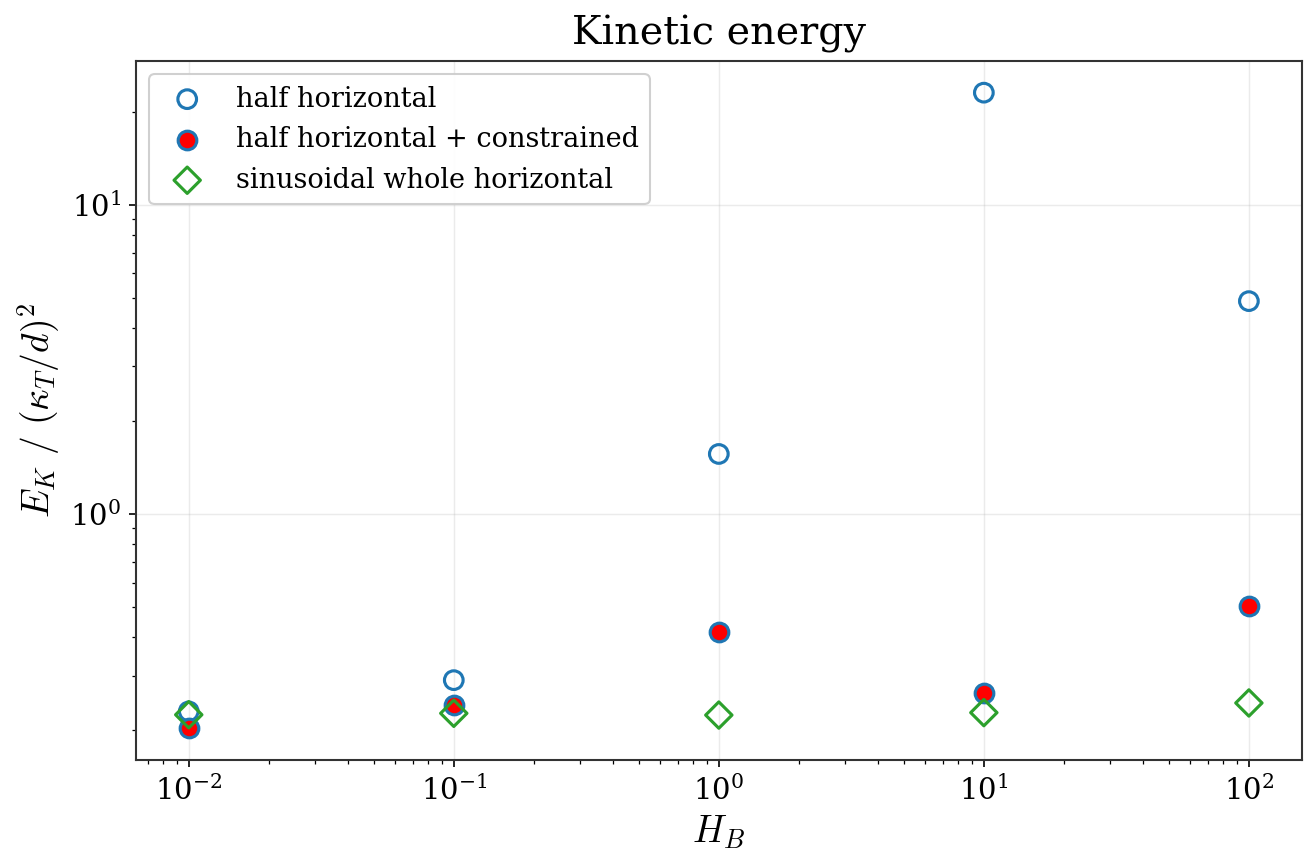}
    \caption{Same as Figure \ref{fig:BE}, but now showing kinetic energy.}
    \label{fig:KE}
\end{figure*}

\begin{figure*}
    \centering
    \includegraphics[width=1\linewidth]{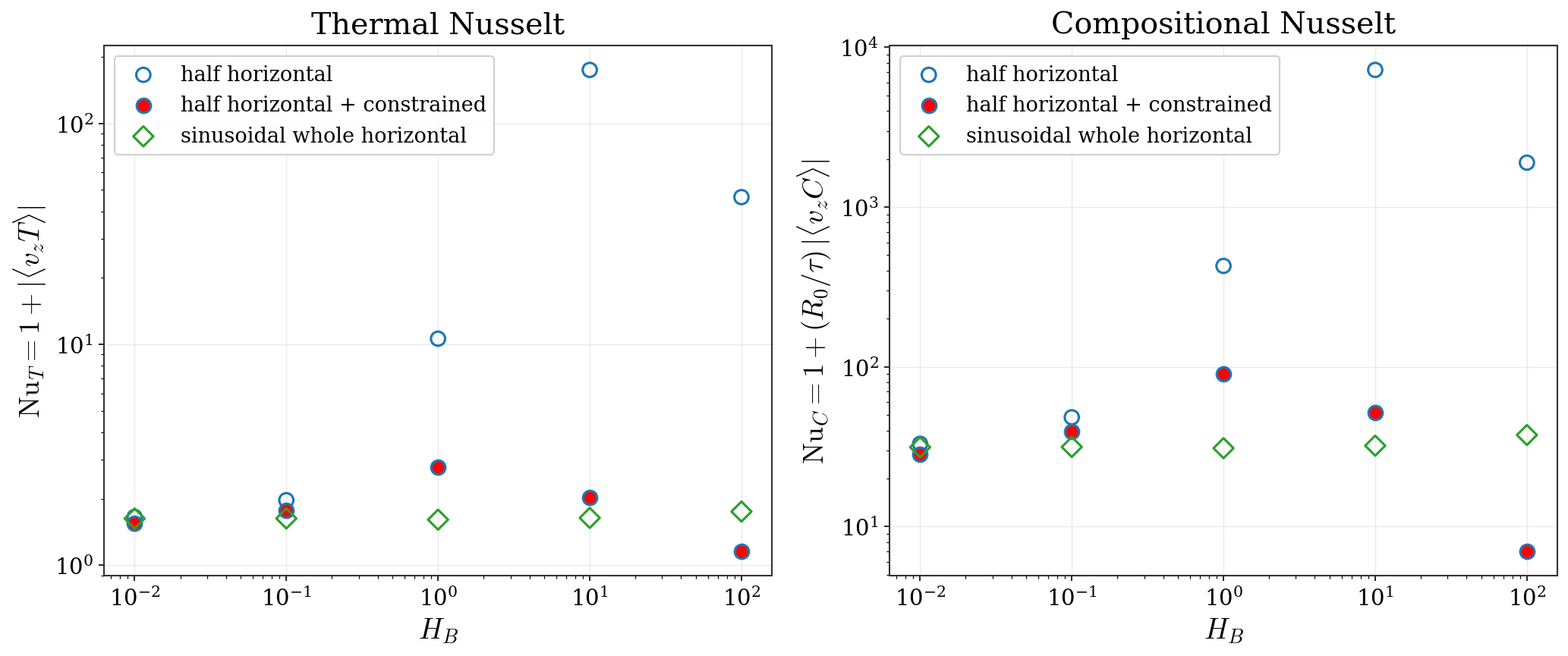}
    \caption{Final volume-averaged and time-averaged composition and temperature Nusselt numbers for the simulations with initial condition given by (Eq. \ref{half_horizontal_initB}) for the circles, and  initial conditions given by (Eq. \ref{sinusoidal_initB}) for the green diamonds over \(H_B\) at $\mathrm{Pm}=1$ . Blue circles with red filling are simulations with no constrain on the \(k_z = 0\) 'elevator' modes, filled blue circles are simulations with constrained 'elevator' modes. }
    \label{fig:nusselt}
\end{figure*}

\section{Discussion}

Our simulations show that magnetic fields of different strengths and initial geometries can change the saturated properties of thermohaline convection. They affect the mean fluid velocity, the resulting magnetic field, the composition flux, and the temperature flux.  
As shown in Figure \ref{fig:KE}, the kinetic energies for the simulations with large initial magnetic flux (half-horizontal and large $H_B$) are orders of magnitude higher due to the presence of elevator modes, and do not reach a steady state. Further, the vertical component of the flow also exhibits intermittency, characterized by quasi-periodic bursts of vertical motion. The temporal behavior of these bursts qualitatively resembles that associated with the tilting instability observed in two-dimensional thermal convection simulations \citep[see, e.g.,][]{Goluskin2014,Fuentes2021}. 
When elevator modes are explicitly forbidden, which is likely more representative of a real star, the kinetic energies can still be larger by a factor of a few. When magnetic fields are in the vertical direction, the enhancement can be much larger \citep{Harrington2019,Fraser:2024}. Therefore, we conclude that magnetic fields generally act to enhance thermohaline turbulent flows.  

Our results suggest that magnetic fields can be transported by thermohaline convection while maintaining magnetic field strengths close to their initial values (Figure \ref{fig:BE}). This is true for both sets of  simulations that start with initially non-zero magnetic flux (half-horizontal simulations, A1-B5). This means that crystallization-induced thermohaline convection may be able to transport magnetic fields from within the core of a white dwarf to outer layers where it could be observed, much faster than the field could diffuse on its own. However, the periodic boundary conditions of our simulations have limited power in understanding realistic systems, because magnetic flux through the boundaries is conserved exactly (unlike a real system) and because elevator-mode-dominated dynamics persist (which cannot occur in a real system). Therefore, performing simulations with more realistic boundary conditions or spherical geometry is a natural next step to determine if magnetic fingering convection can explain the observed magnetic fields of white dwarfs. 

In simulations without any magnetic flux and $\mathrm{Pm}=1$  (whole-sinusoidal simulations, C1-C5) the magnetic energy decays to nearly zero. In this case, the magnetic field is destroyed as it is transported, due to the effective turbulent diffusivity of the thermohaline motions. This may represent what happens in a real star where there is no magnetic helicity, because helicity is approximately conserved in real systems. Hence, a possible extension of our results to real stars is that thermohaline convection can transport magnetic fields outwards while moving towards a minimum magnetic energy that conserves magnetic helicity. In stars with large magnetic helicity, this would allow thermohaline convection to transport the fields to the outer layers where they can be observed.

Our simulations with no net flux and large magnetic Prandtl number, $\mathrm{Pm}=10$, (whole-sinusoidal simulations, D1-D5) are the only ones able to regenerate magnetic fields via a dynamo (Figure \ref{fig:pm10_equipartition}). At low $H_B$, the final magnetic to kinetic energy ratio is $E_B/E_K \approx 0.5$, i.e., the magnetic energy is roughly in equipartition with the kinetic energy. At high $H_B$, the magnetic energy remains at super-equipartition values. However, these simulations may not yet have reached their final equilibrium values (see Figure \ref{fig:pm10_timeseries}), and it is possible that they will eventually settle at near-equipartition fields, just like the low values of $H_B$.

In white dwarfs where thermohaline velocities are limited to $v_{\rm therm} \lesssim 0.1 \, {\rm cm/s}$ \citep[e.g.,][]{Montgomery_Dunlap2024}, equipartition field strengths are $B \sim \sqrt{4 \pi \rho v_{\rm therm}^2}$, which evaluates to $B \lesssim 300 \, {\rm G}$ for $\rho = 10^6 \, {\rm g}/{\rm cm}^3$. It thus appears very unlikely that such dynamo-generated fields can produce the strong fields observed ($B \gtrsim 10^6 \, {\rm G}$) in many magnetic white dwarfs, as explained in the introduction.

Some of our simulations with strong initial fields ($H_B > 1$) develop horizontal zonal flows. We do not completely understand why these flows develop, but they may be related to the effectively 2-dimensional nature of the turbulence that develops, similar to zonal flows that develop in convection constrained by rapid rotation. The strong magnetic fields (in the x-direction) prevent bending of the magnetic field lines (i.e., enforce $k_x \simeq 0$), such that thermohaline motions effectively occur in the $yz$-plane and are uniform in the x-direction. Since fields with $H_B \gg 1$ can very easily be present in white dwarfs (only requiring $B \gtrsim 10^3 \, {\rm G}$), these types of constrained dynamics may be common for thermohaline motion, and should be investigated in future work.

Previous works \cite{Harrington2019, Fraser:2024} have looked at the effects of vertically imposed magnetic fields on thermohaline convection. Our results in different initial field geometries add to their conclusions that magnetic fields present in white dwarfs meaningfully change the energies, velocities, and transport properties of thermohaline convection.  We found that the initial field geometry changes the final state of fully developed fingering convection.
In simulations with horizontal magnetic fields, the combination of elevator modes and magnetic flux conservation through the boundaries create an ongoing intermittent growth state that does not saturate during the time we simulate, and would likely never saturate. The parasitic saturation model proposed by \citet{Brown2013}, extended to incompressible MHD in \cite{Harrington2019} and \cite{Fraser:2024}, does not apply to our simulations with non-zero horizontal magnetic flux and $H_B  > 1$, because the magnetic field strength keeps growing rather than being a fixed quantity. However, both elevator modes and conserved magnetic flux are artifacts of periodic boundary conditions that do not exist in real stars, preventing a straightforward translation of the simulation to realistic systems. Constraining the fastest growing modes allows the simulations to saturate, however, this artificial fix imparts a box size-dependence on the saturated state (see Appendix \ref{convergence}). Therefore, simulations with more realistic geometries and boundary conditions are required to mitigate artifacts associated with triply-periodic boundaries.

\section{Conclusion} \label{sec:end}

We used the Dedalus spectral code to simulate thermohaline convection in 3-dimensional cartesian geometry, using triply periodic boundary conditions. We employed the Boussinesq approximation and included temperature, composition, and magnetic field evolution. We considered cases with initially horizontal magnetic fields that are non-uniform over the simulation domain. The analysis of our solutions shows that the initial geometry of the magnetic field affects the transient regime of the evolution, but in all our simulations with initial magnetic flux, the final state has the magnetic field mostly in the vertical direction while preserving the overall flux.

The unconstrained simulations A4 and A5 are dominated by elevator modes ($k_z=0$) at late times. The z-aligned magnetic fields that develop in these simulations suppress the parasitic shear modes and allow for higher flow velocities to develop. Simulations beginning with a strong magnetic field ($H_B  > 1)$ do not converge in our simulation time and exhibit gradual intermittent growth in their kinetic and magnetic energies. The simulation with the strongest field ($H_B =100$, simulation A5) also shows a long transient dominated by horizontal zonal flows rather than fingers moving in the vertical direction. It is unlikely these results are representative of thermohaline convection in real stars, because the elevator modes and conserved magnetic flux are artifacts of triply periodic boundary conditions.

The simulations in which we constrain the $k_z=0$ modes (simulations B1-B5) all saturate (except for B5, which also develops zonal flows). For weak initial fields, ($H_B \leq 1$), the saturated magnetic fields  are a few times higher than their minimum and their initial field energy due to generation of vertical magnetic fields by the thermohaline fingers. For strong initial fields ($H_B  > 1$), magnetic field energies saturate close to their minimum possible values. 

All our simulations (A1-D5) show averaged Nusselt numbers ($\text{Nu}_T$ and $\text{Nu}_C$) greater than unity (Figure \ref{fig:nusselt}). This implies that the magnetic fingering convection is more efficient at transporting temperature and composition than regular diffusion. However, we note that the saturated states of the B1-B5 simulations depend on the size of the simulation domain, so the kinetic energy and Nusselt numbers realized by a more realistic (larger) domain are unclear; see further discussion in Appendix \ref{convergence}.  

Simulations with zero initial flux C1-D5 have two distinct regimes based on $\mathrm{Pm}$. Simulations C1-C5 have lower magnetic Prandtl number ($\mathrm{Pm}=1$, larger magnetic diffusion) and the magnetic field decays via ohmic diffusion over time. The saturated state kinetic energies are almost the same and the magnetic fields are negligible. Simulations D1-D5 have a higher magnetic Prandtl number ($\mathrm{Pm}=10$, smaller magnetic diffusion) allowing for a small scale dynamo that enhances the field close to equipartition with the kinetic energy (Figure \ref{fig:pm10_timeseries}).  In our MESA models,   $\mathrm{Pm} \approx 10^{-1}-10^2$ so the small scale dynamo may be present in WD thermohaline regions.

Applied to white dwarfs, our simulations suggest only weak magnetic fields can be generated by thermohaline motions, and only in cases where $\mathrm{Pm}>1$. However, our simulations tentatively indicate that strong magnetic fields can be advected outwards by thermohaline motion without being destroyed. The conserved magnetic flux in our simulations make it impossible to reach a robust conclusion, but we speculate that realistic systems would transport magnetic fields outwards while approximately conserving magnetic helicity. Future work with realistic stellar geometry should be performed to better understand the interplay of thermohaline motion and magnetic fields in crystallizing white dwarfs.

\begin{acknowledgments}
The authors thank Daniel Lecoanet for answering questions about implementations of our simulations in Dedalus and helpful discussions. We thank Adrian Fraser for suggesting testing the box height dependence of our constrained simulations and pointing out that MHD thermohaline convection can exhibit small scale dynamos. This work was partially funded by support from the United States–Israel Binational Science Foundation through grant BSF-2022175. The simulations shown in this work were performed in Caltech HPC and in Purdue Anvil supercomputer through the NSF ACCESS project PHY250199. J.R.F. is supported by the Sherman Fairchild Postdoctoral (Burke) Fellowship and the Presidential Fellowship at Caltech, as well as NASA Solar System Workings grant 80NSSC24K0927.
\end{acknowledgments}

\appendix

\section{MESA models and dimensionless numbers}
\label{MESA_models}

We used the Modules for Experiments in Astrophysics (MESA) \cite{MESA2013,MESA2018,Bauer_2026} code to create models of carbon-oxygen (C/O) white dwarfs with masses ranging from 0.5  \(M _\odot\) to 1.0 \(M _\odot\). The  module we used was developed in \citet{Bauer_2026} and contains the essential physics for our description: thermohaline mixing prescriptions based on the Brown model or the Kippenhahn model. In addition, it includes crystallization physics and C/O separation prescription. This model captures the physics that we want to model in our direct numerical simulation. 

Following works by \citet{Garaud2015,NandkumarPethick1984}, we calculated the transport coefficients corresponding to the liquid region of C/O white dwarfs using the evolved MESA profiles. The calculated transport coefficients are the thermal \(\kappa_T\), viscous \(\nu\), magnetic \(\eta\), and chemical diffusivities \(\kappa_C\). The Brunt-Väisälä frequencies of the composition and thermal profiles were also calculated.  

The thermal diffusivity is calculated using only its radiative contribution, but inputting the effective opacity calculated by our MESA model

\begin{equation*}
    \kappa_T = \frac{16 \sigma_{\rm SB} T^3}{3 \kappa_{\text{eff}} \rho^2 c_p}
\end{equation*}

\noindent where $\rho$ is density, $\sigma_{\rm SB}$ is the Stefan-Boltzmann constant, $\kappa_\text{eff}$ is the effective opacity and accounts for the effect of radiative diffusion and electron conduction, and $c_p$ is the heat capacity at constant pressure.

The viscosity is a combination of three components, the radiative viscosity $\nu_\text{rad}$, the ionic viscosity $\nu_\text{ion}$ taken from \citet{Garaud2015}, and the electronic viscosity $\nu_\text{e}^\text{NP}$ calculated using the semi-analytic viscosity estimate of Nandkumar and Pethick, which is the main contribution in our strongly degenerate fluid (see Eq. 16 of \cite{NandkumarPethick1984}):

\begin{align*}
    \nu &= \nu_\text{rad} + \nu_\text{ion} + \nu_\text{e}^\text{NP} \nonumber \\
    =&  \frac{16 \sigma_{SB} T^4}{15 c^2 \kappa \rho^2} + \frac{0.4066 m_H^{1/2} (k_B T)^{5/3}}{e^4 C(\ln \Lambda_{CO})\rho} +\nu_\text{e}^\text{NP}
\end{align*}

\noindent where $e$ is the fundamental charge, $m_H$ is the mass of a proton, c is the speed of light,  $\kappa$ is the opacity, $k_B$ is the Boltzmann constant, and $\ln \Lambda_{CO}$ is the coulomb logarithm with carbon and oxygen. 
\begin{figure*}
    \includegraphics[width=1\linewidth]{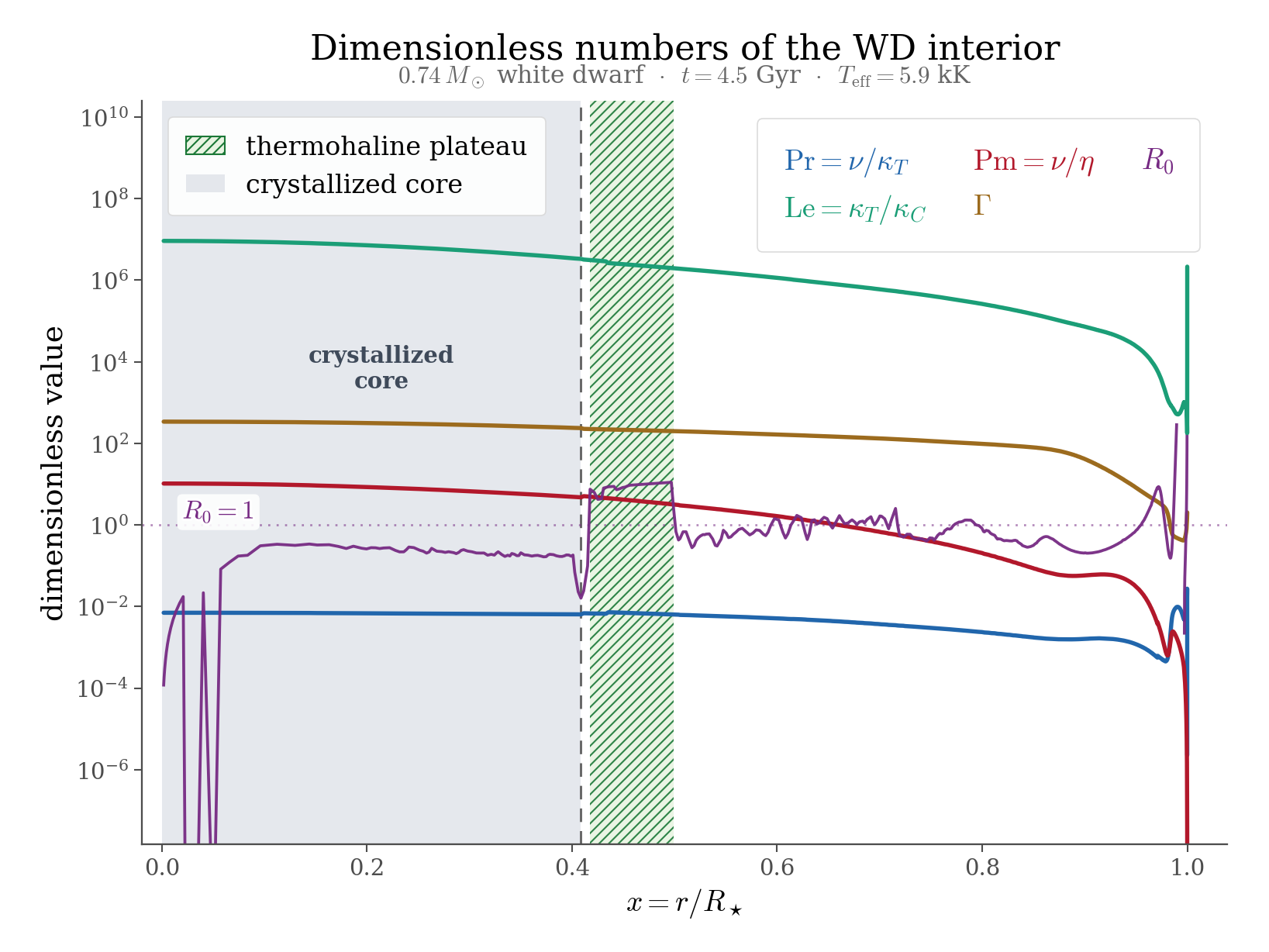}
     
    \caption{Fluid parameters as a function of radius in a $M=0.74 \, M_\odot$ WD model at $t=4.5$ Gyr and $T_{\text{eff}}=5900 K$. The green shaded region labeled "thermohaline plateau" above the core satisfies the condition $1<R_0<\mathrm{Le}$, and is where we expect thermohaline instability to occur within this WD. }
    \label{fig:models}
\end{figure*}

\begin{figure*}
    
     \includegraphics[width=1\linewidth]{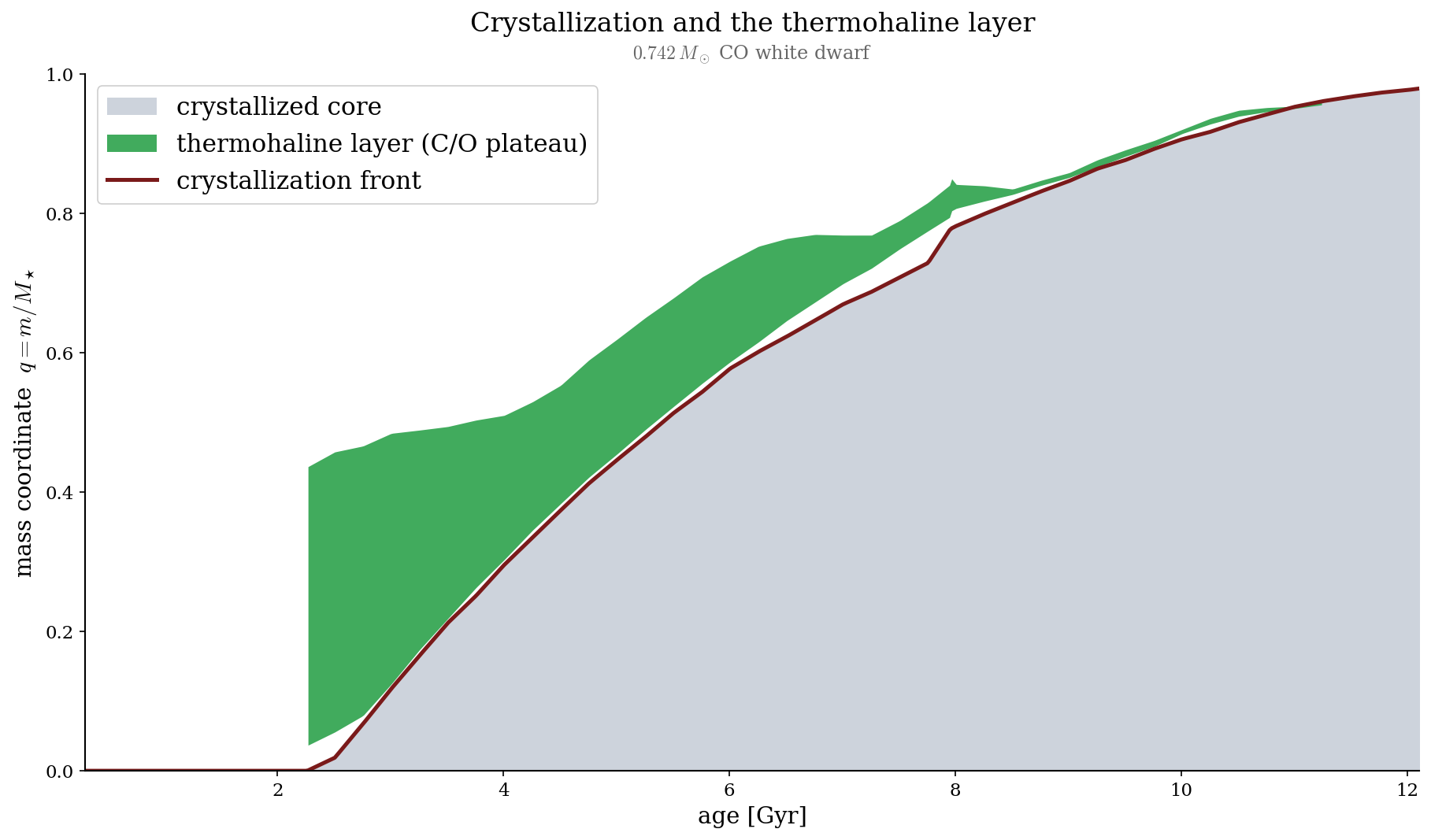}
    \caption{Kippenhahn diagram showing the crystallization of a 0.74 \(M_\odot\) C/O white dwarf. The green region shows the thermohaline layer above the crystallization front (red line) atop the crystalline core (gray region).}
    \label{fig:Kippenhahn-diagram}
\end{figure*}
The electronic viscosity calculated in \citet{NandkumarPethick1984} assumes the Coulomb parameter of $\Gamma \leq 175$ (This parameter is the ratio of the mean Coulomb potential of the ions over mean thermal energy) and the region exhibiting thermohaline has $\Gamma > 200$ (see Fig. \ref{fig:models}). This strong coupling regime of the ions affects the ion-electron component of the electronic viscosity. Nevertheless, taking into account this coupling dependence in the calculation of $\nu_e$, as done in \citet{Chugunov2005Shear}, viscosity is reduced only by a factor of order $O(1)$ in the thermohaline regions of the MESA models considered.

The compositional diffusivity is taken from Eq. (30) of \citet{Garaud2015} based on the prescription by \citet{Michaud1993}

\begin{equation}
    \kappa_{\mu}= \frac{15}{16 \rho \ln \Lambda_{\mathrm{CO}}} \sqrt{\frac{2 m_H}{5\pi}} \frac{(k_B T)^{5/2}}{e^4} \frac{3+X}{(1+X)(3+5X)(0.7+0.3X)} \; .
\end{equation}
where $X$ is the hydrogen fraction that we set to 0. Applying this diffusivity to strongly coupled dense ion mixtures without hydrogen is outside of its regime of applicability, but we expect this diffusivity to be very small in our system so this crude estimate is enough for our purposes.

MESA computes the magnetic diffusivity $\eta = c^2 /4 \pi \sigma$ from the three-regime fit of \citet{Spitzer1965, Wendell1987, NandkumarPethick1984} implemented by \citet{Yoon2004}. Since the liquid region is highly degenerate, all the numbers we quote are calculated in the regime of \citet{NandkumarPethick1984}. The same caveat applies to the $\Gamma>200$ strong coupling regime of the Coulomb fluid. Using the prescription of \citet{Potekhin_2000} that takes this coupling into account, the factor difference also remains $O(1)$ in the thermohaline region.

The transport coefficients and Brunt-Väisälä frequencies enter our numerical scheme in the definitions of dimensionless numbers described in Section \ref{Boussinesqequations}. We prioritized setting the correct ordering of dimensionless numbers based on transport properties rather than trying to match any quantity precisely. Using the quantities in the MESA models and the expressions for the transport coefficients above, we find that the dimensionless parameters in the unstable thermohaline region of the interiors of WD are of the order of $\mathrm{Pr}  \approx 10^{-2} $, $\mathrm{Pm} \approx 10^{-1}-10^2$, $\mathrm{Le}  = 10^6 - 10^8$, $R_0  \approx 10^0 - 10^2$  (see  Figure \ref{fig:models} for the example of $0.74 M_{\odot}$).  Additionally, we show that in our models there is a thermohaline unstable region above the crystalline core. In the Kippenhahn diagram (see Figure \ref{fig:Kippenhahn-diagram}), we estimate the size of the thermohaline unstable region above the crystallizing core using the radial size of the plateau of C/O element fractions above the crystal front. The region where thermohaline convection is active has nearly uniform fractions of C and O due to efficient thermohaline mixing.

\section{Flux conservation and minimum field}
\label{fluxConservationProof}

In a triply periodic cartesian 3D box, the magnetic flux through any side of the box is conserved. We can see this by considering the induction equation~\eqref{eq:B}. Integrating both sides over a surface S that has boundaries on the sides of the box gives us
\begin{equation}
\int_S \frac{\partial \bm{B}}{\partial t} \cdot dA = \frac{dQ}{dt}  = \int_S \nabla \times (\bm{v} \times \bm{B}) \cdot dA + \int_S \eta \nabla^2 \bm{B} \cdot dA \, ,
\end{equation}
where $Q$ is the magnetic flux through the surface.

Using the identity \( \nabla \times (\nabla \times \bm{B}) = \nabla (\nabla \cdot \bm{B}) - \nabla^2 \bm{B}\) and the solenoidal character of the magnetic field \(\nabla \cdot \bm{B} = 0\)
\begin{equation}
    \frac{dQ}{dt} = \int_S \nabla \times (\bm{v} \times \mathbf{B}) \cdot dA - \int_S  \eta \nabla \times (\nabla \times \bm{B}) \cdot dA \, .
\end{equation}
We can use Stokes theorem to simplify the integrals in the right-hand side
\begin{equation}
\frac{dQ}{dt} = \int_{\delta S} (\bm{v} \times \bm{B}) \cdot dl + \int_{\delta S} \eta \nabla \times \bm{B} \cdot dl
\end{equation}
where the integral is now along the boundary of the surface. For periodic boundaries, every line element of this integral has a contribution where \( dl \rightarrow -dl\) on the other side of the surface, where the line integral traverses the same location in the opposite direction. Therefore, the right hand side always integrates to zero. Thus, 
\begin{equation}
    \frac{dQ}{dt} = 0 \,
\end{equation}
and so the magnetic flux is conserved exactly.

Now, we will show that the conservation of magnetic flux through the boundaries of the domain leads to a conserved lower bound on the total magnetic energy $E_B$. To start, let us assume our magnetic field $B$ starts perfectly horizontal in the x-direction with its magnitude depending on the z-coordinate $B = b(z) \hat{x}$.

Then, we consider the magnetic flux through a surface S that we take to be the $yz$ plane at $x = 0$. 

\begin{equation}
Q = \int^{\frac{L}{2}}_{-\frac{L}{2}} \int^{\frac{L}{2}}_{-\frac{L}{2}} b(z) \,dy\,dz =  L \int^{\frac{L}{2}}_{-\frac{L}{2}} b(z)\,dz
\end{equation}

Using the initial condition (Eq. \ref{half_horizontal_initB}), in the limit of $\epsilon \rightarrow \infty$, our $b(z)$ can be written as 

\begin{align}
        b(z) = \begin{cases}
                -\sin\left( \frac{2 \pi}{L} z \right), & -\frac{L}{2} \leq z \leq 0, \\
                0, & 0 \leq z \leq \frac{L}{2}.\\
               \end{cases}
    \end{align}

Now, we can calculate the flux through the surface S 

\begin{equation}
    Q = -L \int^0_{-\frac{L}{2}} \sin \left( \frac{2 \pi }{L} z \right) dz = \frac{L^2}{2 \pi} \cos \left( \frac{2 \pi}{L}   z \right) \Big|^{0}_{-\frac{L}{2}} = \frac{L^2}{\pi} > 0
\end{equation}

Using the flux conservation that we proved earlier, we can bound the magnetic energy in the x direction from below where $B(t,\textbf{x}) = (B_x (t,\textbf{x}),B_y (t,\textbf{x}),B_z (t,\textbf{x}))$

\begin{align}
    Q = \int_S B_x(t, \textbf{x}) dy dz \\
    |Q| \leq \int_S |B_x(t, \textbf{x})| dy dz  = |B_x| L^2 \\
    \implies \frac{1}{\pi} \leq |B_x| \quad \text{thus,} \quad \frac{H_B}{2 \pi^2} \leq \frac{H_B}{2}\bm{B} \cdot \bm{B} = \langle E_B \rangle
\end{align}

Here \(\langle E_B \rangle\) is the magnetic energy averaged over the volume. Therefore, our simulations that start with non-zero initial magnetic flux have lower bound magnetic energy throughout the whole evolution of the simulation.

\section{Convergence analysis}
\label{convergence}

\begin{figure*}
    \centering
    \includegraphics[width=1\linewidth]{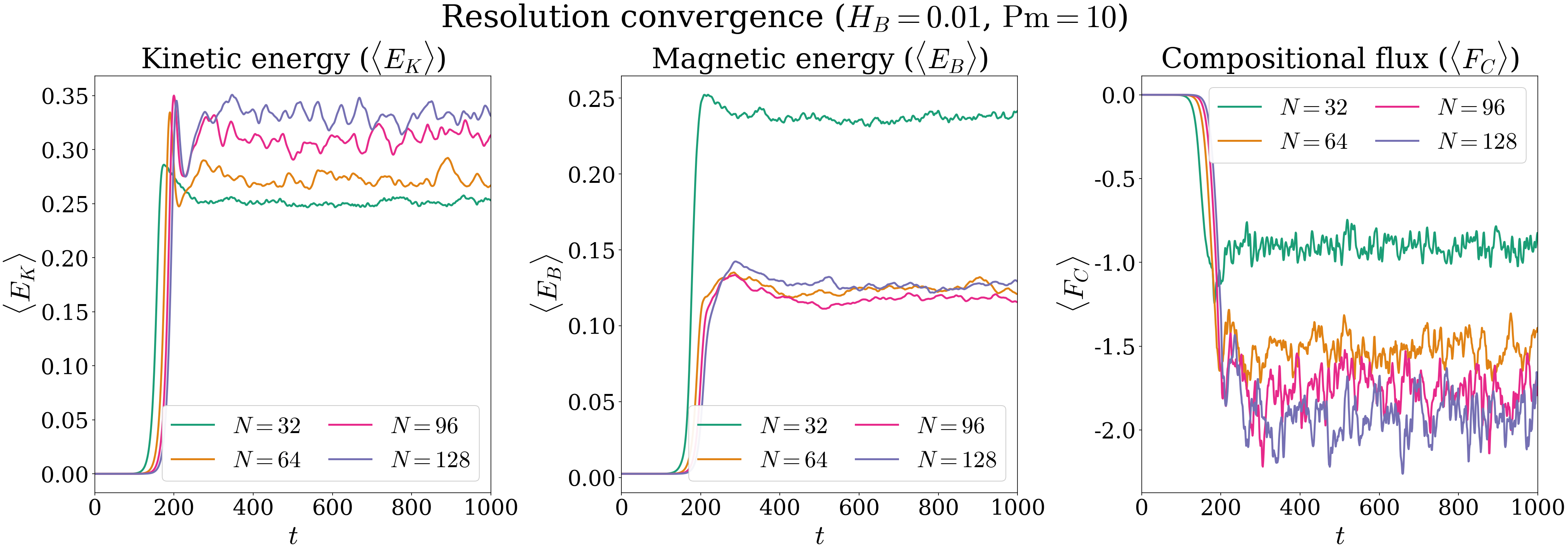}
    \caption{The volume average of the  kinetic energy, magnetic energy and composition flux from left to right of simulations with different resolutions N=32,64,96,128. These simulations have whole-sinusoidal initial conditions of Eq. ~\eqref{sinusoidal_initB} and dimensionless numbers $\mathrm{Pr} =0.1$, $\mathrm{Le}  = 10$, $\mathrm{Pm}  = 10$, $H_B  = 0.01$, $R_0 =3$.}
    \label{fig:convergenceHb0.01}
\end{figure*}

We tested our simulations for convergence with respect to the simulation box size. We ran simulations with the whole-sinusoidal initial condition for resolutions $N=32,64,96,128$ and plot the volume-averaged kinetic energy $ E_{K}$, the volume-averaged magnetic energy $ E_{B} $, and the volume-averaged compositional flux $F_C$. We see in Figure \ref{fig:convergenceHb0.01}
that the three quantities seem to converge as the resolution increases. The $N=96$ and $N=128$ cases are almost identical in the early growth stage of the evolution and only differ slightly in the latter relaxation stage and equilibrium. This also confirms that the small scale dynamo we observe in our $\mathrm{Pm}  = 10$ simulations is numerically robust. 

\begin{figure*}
    \centering
    \includegraphics[width=1\linewidth]{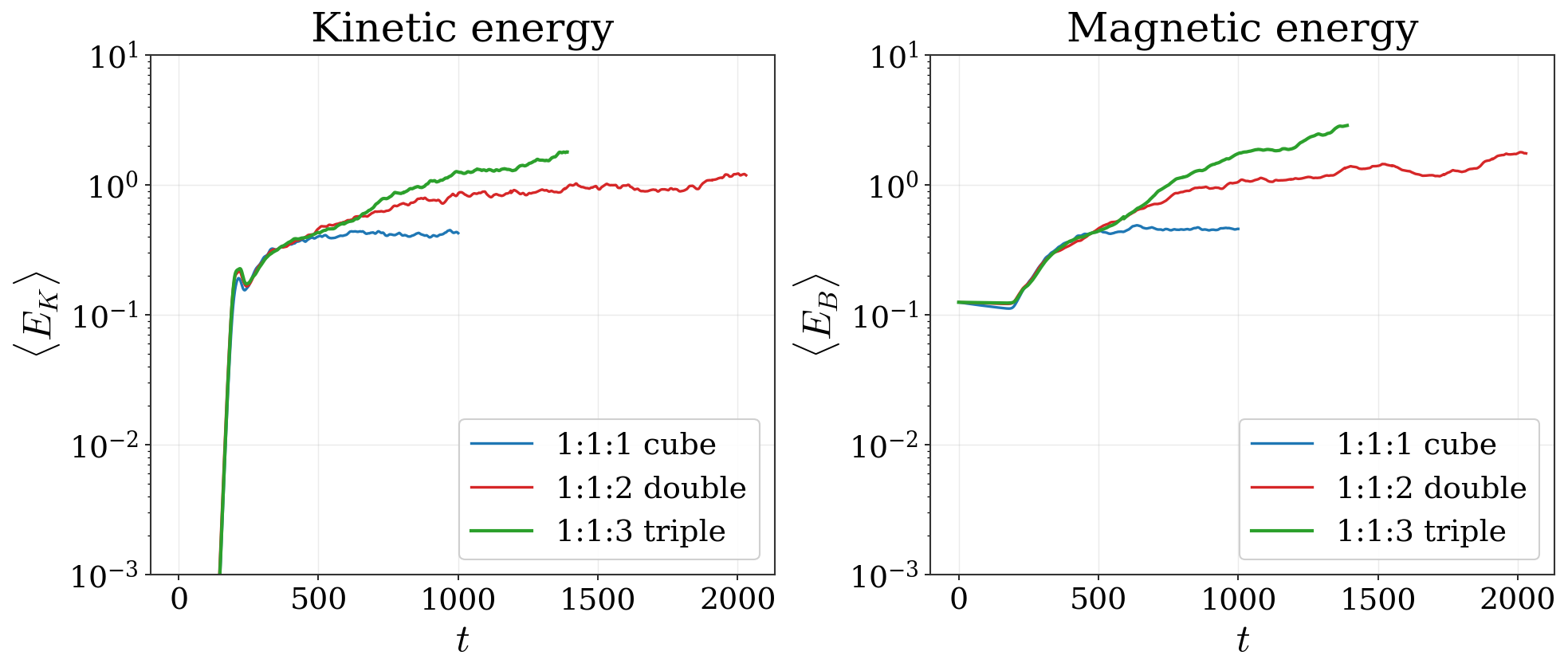}
    \caption{Energies for simulations with constrained elevator modes and dimensionless numbers $\mathrm{Pr} =0.1$, $\mathrm{Le}  = 10$, $\mathrm{Pm}  = 1$, $H_B  = 1$, $R_0 =3$. We see that both magnetic and kinetic energies are dependent on the simulation domain size $L_z$  in the constrained elevator mode simulations.}
    \label{fig:boxsize}
\end{figure*}

We perform a second consistency test on constrained half-horizontal simulations with $\mathrm{H_B} =1$, $\mathrm{Pm}=1$, $R_0 =3$, $\mathrm{Le} =10$ and with $L_z=100d,200d,300d$ which we refer to as the cube 1:1:1 (B3), double 1:1:2 (F1), and triple 1:1:3 (F2) simulations, respectively. The energies of the cube (blue curve), double (red curve), and triple (green curve) simulations are shown in Figure \ref{fig:boxsize}. The cube simulation appears to have reached saturation, but double and triple have not converged in our simulated time. We believe that this box size dependence, for constrained \(k_z=0\) modes, occurs because when we increase \(L_z\) smaller \(k_z\) modes become available, and they start to dominate the dynamics as the fastest growing modes. In the \(L_z \rightarrow \infty\) limit we expect the \(k_z \rightarrow 0\) modes to be the fastest growing modes, but we also expect parasitic modes to eventually saturate the flow in that limit. When no constraint is applied to the elevator modes, the averaged energies are independent of the box size. 

\section{Additional field geometries}

\subsection{Gaussian noise field initial condition}

\begin{figure*}

    \includegraphics[width=1\linewidth]{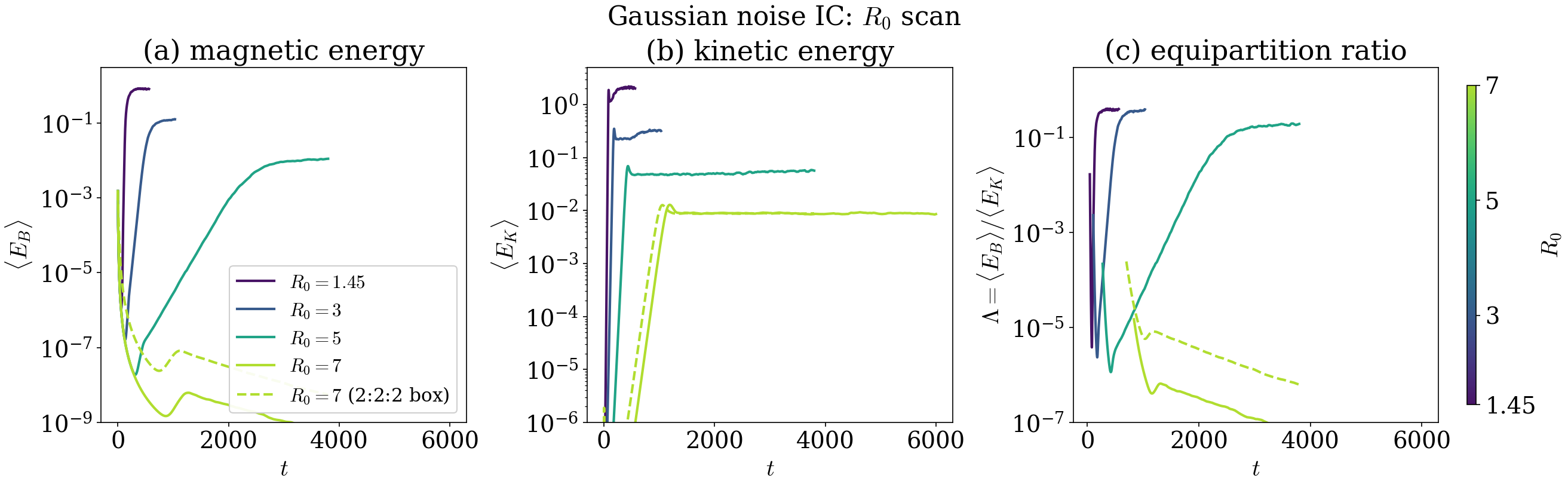}
    \caption{Thermohaline convection and small-scale dynamo volume-averaged quantities for simulations with small random initial magnetic fields. The simulation parameters are displayed in Table  \ref{tab:summarytable} for E1-E5 simulations. Decreasing $R_0$ causes more vigorous convection that translates to a more intense dynamo. The small-scale dynamo failed to manifest in both $R_0 =7$ simulations E4 and E5.}
    \label{fig:gnoise}
\end{figure*}

In order to separate the effect of a small scale dynamo (SSD) from the properties of the initial field, we performed simulations with Gaussian noise field initial conditions. The magnetic fields in these simulations were initialized by adding Gaussian noise of mean amplitude $O(10^{-3})$ into our units. We also varied the parameter $R_0$ for these runs (E1-E5). These simulations show that the small scale dynamo present in D1-D5 and E1-E5 saturates with $E_B/E_K \approx 0.35-0.4$ and that smaller $R_0$ translates to both enhanced fingering convection and enhanced dynamo action (see Fig. \ref{fig:gnoise}). Additionally, the simulations at $R_0 = 7$ did not exhibit a small-scale dynamo because it has less vigorous convection, so the magnetic Reynolds number $Re_m$, proportional to $u_\text{rms}$, was too small to trigger the dynamo.

\bibliographystyle{aasjournal}
\bibliography{references}

\end{document}